\documentclass[aps,pra,superscriptaddress,reprint,a4paper]{revtex4-2}
\usepackage{subcaption}
\usepackage{ragged2e}
\usepackage[english]{babel}
\usepackage[letterpaper,top=2cm,bottom=2cm,left=3cm,right=3cm,marginparwidth=2cm]{geometry}

\usepackage{multirow}
\usepackage{graphicx}
\usepackage[draft]{hyperref}
\usepackage{float}
\usepackage{amsmath,amsfonts}
\usepackage{amsthm}
\usepackage{amssymb}
\usepackage{bm}
\usepackage{placeins}
\usepackage{orcidlink}
\usepackage[normalem]{ulem}

\usepackage{xcolor}
\usepackage[colorinlistoftodos]{todonotes}
\usepackage{tabularx}
\usepackage{mathtools}
\newcommand{\mbf}[1]{\mathbf{#1}}
\newcommand{\mrm}[1]{\mathrm{#1}}

\newcommand{\pr}{\mathrm{PR}}

\newcommand{\ii}{\mathrm{i}}
\newcommand{\ket}[1]{\left| #1 \right\rangle }
\newcommand{\bra}[1]{\left\langle #1 \right| }
\newcommand{\ketbra}[2]{\ket {#1}\!\! \bra {#2} }

\newcommand{\eqnref}[1]{Eq.~(\ref{#1})}

\newcommand{\ww}{\omega}

\newcommand{\rmdn}[2]{#1_{\mrm{#2}}}

\newcommand{\ehat}{\boldsymbol{\hat{\epsilon}}}

\newcommand\scalemath[2]{\scalebox{#1}{\mbox{\ensuremath{\displaystyle #2}}}}

\newcommand{\wRF}{\omega_{\mathrm{rf}}}

\begin{document}

\title{Optimized detection modality for double resonance alignment based optical magnetometer}
\author{A.~Akbar\orcidlink{0000-0001-9915-6117}}
\email{ali.akbar@nottingham.ac.uk}
\affiliation{School of Physics and Astronomy, University of Nottingham, University Park, Nottingham, NG7 2RD, UK}

\author{M.~Ko\'{z}bia\l\orcidlink{0000-0002-2709-6677}}
\affiliation{Centre for Quantum Optical Technologies, Centre of New Technologies, University of Warsaw, Banacha 2c, 02-097 Warszawa, Poland}

\author{L.~Elson\orcidlink{0000-0002-0863-6447}}
\affiliation{School of Physics and Astronomy, University of Nottingham, University Park, Nottingham, NG7 2RD, UK}

\author{A.~Meraki\orcidlink{0000-0002-0780-5034}}

\affiliation{School of Physics and Astronomy, University of Nottingham, University Park, Nottingham, NG7 2RD, UK}

\author{J.~Ko\l{}ody\'{n}ski\orcidlink{0000-0001-8211-0016}}

\affiliation{Centre for Quantum Optical Technologies, Centre of New Technologies, University of Warsaw, Banacha 2c, 02-097 Warszawa, Poland}

\author{K.~Jensen\orcidlink{0000-0002-8417-4328}} 
\email{kasjensendk@gmail.com}
\affiliation{School of Physics and Astronomy, University of Nottingham, University Park, Nottingham, NG7 2RD, UK}

\begin{abstract}
In this work, we present a comprehensive and comparative analysis of two detection modalities, i.e., polarization rotation and absorption measurement of light, for a double resonance alignment based optical magnetometer (DRAM). We derive algebraic expressions for magnetometry signals based on multipole moments description.
Experiments are carried out using a room-temperature paraffin-coated Caesium vapour cell and measuring either the polarization rotation or absorption of the transmitted laser light.
A detailed experimental analysis of the resonance spectra is performed to validate the theoretical findings for various input parameters. 
The results signify the use of a single isotropic relaxation rate thus simplifying the data analysis for optimization of the DRAM. The sensitivity measurements are performed and reveal that the polarization rotation detection mode yields larger signals and better sensitivity than absorption measurement of light.  
\end{abstract}

\maketitle
\section{Introduction}
Optically pumped magnetometers (OPMs) \cite{budker2007optical,budker2013optical,benumof1965optical} based on alkali atom vapours have been instrumental in the advancement of diverse fields including magnetoencephalography (MEG) \cite{xia2006magnetoencephalography,sander2012magnetoencephalography,borna201720,iivanainen2019scalp,brookes2022magnetoencephalography}, biomagnetism\cite{jensen2018magnetocardiography,jensen2022biomag,Chalupczak2012,kowalczyk2021detection},  paleomagnetism \cite{romalis2011atomic}, remote object detection  \cite{rushton2022unshielded,deans2018active} and dark matter research \cite{masia2020analysis,pospelov2013detecting}. The recent increase in the scope of OPMs can be attributed to their non-destructive, remote sensing capabilities to measure weak magnetic fields in the range of fT/$\sqrt{\text{Hz}}$ \cite{zhao2023optically} which were possible earlier only with superconducting quantum interference device (SQUID) magnetometers operating at cryogenic temperatures \cite{granata2016nano}. The technological advancement in the field of optical magnetometry has led to the enhanced sensitivity (sub-fT/$\sqrt{\text{Hz}}$), miniaturization, robust wearable capabilities of the OPMs and has potentially opened up new avenues in medical science to be explored \cite{clancy2021study,limes2020portable,rhodes2022turning,hillebrand2023non}. 

OPMs have been investigated as a double resonance optical magnetometer \cite{Weis2016bookchapter} which can be used as a scalar magnetometer for e.g.~measuring the Earth's magnetic field amplitude, or for detecting oscillating/radio-frequency (RF) magnetic fields \cite{bevington2019enhanced,bevington2020object}.
In a double resonance optical magnetometer, light is used for polarizing the atomic spins using the process of optical pumping. The spins are then driven by a RF-field $\mathbf{B}_{\text{rf}}$(t) creating precession of the spin-polarization around a static magnetic field $\mathbf{B}_{0}$.
In a double resonance orientation magnetometer (DROM), circular polarized light is used to orient the atomic spins in a certain direction. On the other hand, in a double resonance alignment magnetometer (DRAM), linear polarized light is used to align the atomic spins along a certain axis \cite{weis2006theory}.
In both cases, the spin-precession will induce periodic modulations in the measured observables, i.e., the absorption or polarization rotation of the transmitted light. These modulations will in turn generate signals at first and/or second harmonic of the applied RF frequency. The alignment based magnetometer has been utilized previously for detection of RF fields \cite{di2006experimental,di2007sensitivity}, for zero-field magnetometry \cite{meraki2023zerofield}, vector magnetometry \cite{ingleby2018vector} and also for spin noise spectroscopy \cite{kozbial2023spin}.

The theoretical description of DROM and DRAM for the evolution of spin polarized system is based on a three step model: pump, evolve and probe.
The model is valid for arbitrary RF intensities at low light power, i.e., optical pumping rate is smaller than the relaxation rate such that the system reaches a steady state. Previous works has mainly considered probing of a DRAM through absorption measurement of the transmitted light \cite{weis2006theory}. 

In this work, we extend the theoretical approach for DRAM to another detection modality, i.e., the measurement of the polarization rotation of the transmitted light. We derive the algebraic expressions characterizing the in-phase and quadrature lineshapes where magnetic resonance spectra is observed through rotation of the linear polarization of light (Sec.~\ref{sec;polarization theory}).  Consequently, we performed detailed experimental investigations to explore the dependence of magnetic resonance signal on various parameters to validate the theoretical findings. The detailed analysis for RF intensity, light power and angular dependence of resonance spectra is presented in Section~\ref{sec;results and discussion}. Furthermore, magnetic field sensitivity measurements are performed and results are presented in Section~\ref{sec;sensitivity measurement}. We also present the results for absorption measurements in parallel to draw a comprehensive comparison between the two detection techniques. This work highlights the advantage of utilizing the polarization based detection mode. The results demonstrate that the polarization rotation-based detection outperforms the conventional absorption measurements in terms of magnetic field sensitivity within the DRAM geometry. 
\section{Theoretical Description of DRAM Geometry}
\begin{figure}
    \centering
    \includegraphics[width=0.49\textwidth]{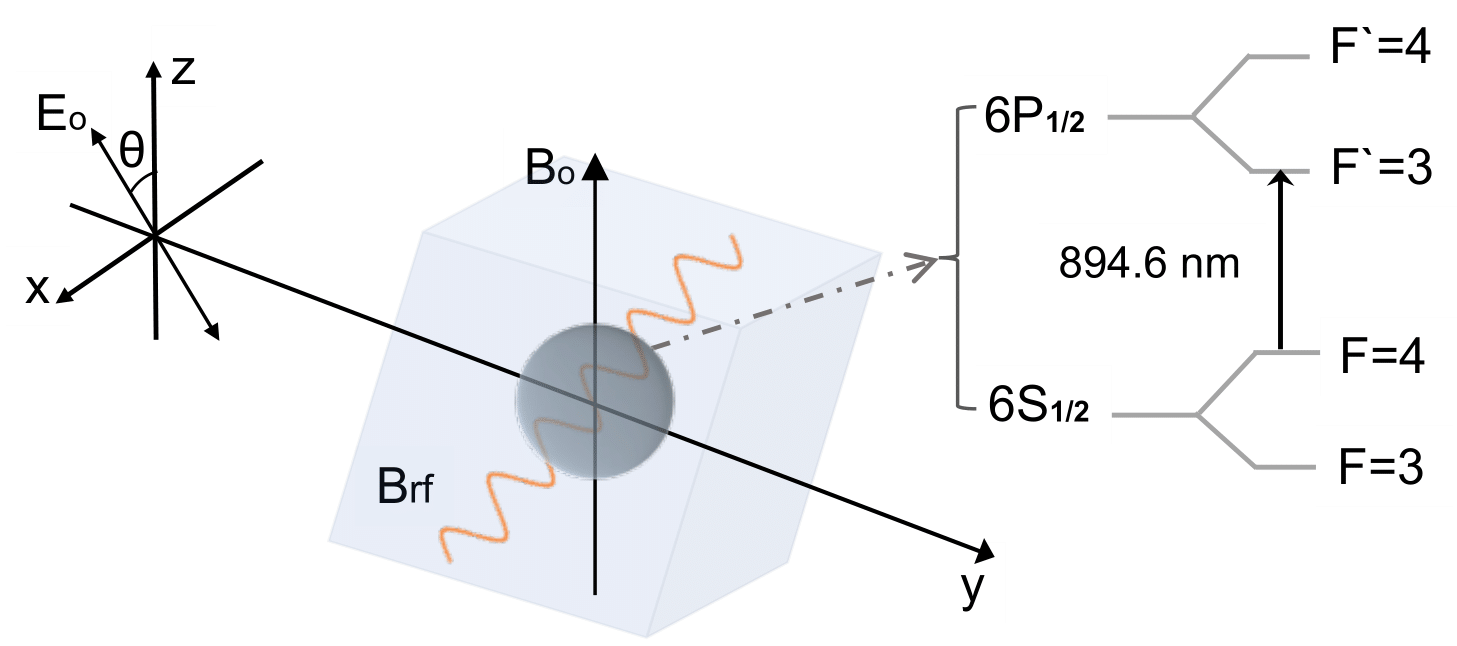}
    \caption{\RaggedRight (a) Illustration of the DRAM geometry where linearly polarized light interacts with Caesium (Cs) atomic vapours in the presence of a static and oscillating magnetic field. (b) Caesium level scheme. The laser light is tuned to the D$1=894.6$ nm transition. 
    } 
    \label{fig:DRAM illustration}
\end{figure}
\subsection{DRAM geometry}\label{sec;polarization theory}
Fig.~\ref{fig:DRAM illustration} illustrates the DRAM geometry. Laser light propagating in the $\mathbf{k} || \hat{y} $ direction and linearly polarized in the $\ehat=\cos(\theta) \mathbf{\hat{z}} + \sin(\theta) \mathbf{\hat{x}}$ direction at an angle $\theta$ in the $zx$-plane is incident on an atomic vapour cell. The linearly polarized light creates a spin-alignment of  atomic vapour which evolves in the combined presence of a static $\mathbf{B}_0$ and an oscillating weak magnetic field $\mathbf{B}_{\text{rf}}(t)$ such that the spin-aligned media will precess around the static magnetic field. The atomic state can be detected by measuring the absorption or the change in polarization of the transmitted light. The signal will oscillate at frequencies at the first and second harmonics of the RF field.

\subsection{Description of the atomic state}
Our experiments are carried out using room-temperature Caesium (Cs) atomic vapour contained in a 5~mm cubic paraffin-coated glass cell.
Caesium has two ground states with hyperfine quantum numbers $F=3$ and $F=4$ and first excited states with $F'=3$ and $F'=4$ as shown in Fig.~\ref{fig:DRAM illustration}. The laser light has wavelength 894.6~nm and is resonant with the $F=4 \rightarrow F'=3$ D1 transition. We consider atoms in the $F=4$ ground state which has $2F+1$ magnetic sub-levels $\ket{F,M}$ (using the $z$-axis as quantization direction) as only those atoms are probed by the light.
A single atom with the hyperfine quantum number $F$ can be described by the density matrix with eigenbasis $\hat{F}^2,\hat{F}_z$ and can be written as
\begin{equation}
    \rho=\sum_{M,M'=-F}^{F}\rho_{MM'}^{(F)}\ketbra {F,M\phantom{'}}{F,M'}.
\label{eq:rho_FM}
\end{equation}
It is instructive to write the density matrix in the basis of spherical tensor operators
\begin{equation}
    \rho=\sum_{\kappa=0}^{2F}\sum_{q=-\kappa}^{\kappa}m_{\kappa,q} \tau^{(\kappa)}_q, 
    \label{eq:rho_m}
\end{equation}
where $m_{\kappa,q}$ are multipole components with rank $\kappa=0,1,...,2F$ and $q=-\kappa,...,\kappa$  and $\tau_q^{\kappa}$ are irreducible spherical tensor operators \cite{weis2006theory, Auzinsh2010book}. For $\kappa=1$, multipoles describe the orientation of the spin whereas $\kappa=2$ describes how the spins are aligned along some axis. We can organize the rank 2 alignment multipoles in a column vector
\begin{equation}
\mbf{m}_2 = \left(m_{2,-2}, m_{2,-1}, m_{2,0}, m_{2,1}, m_{2,2} \right)^T,
\end{equation}  
where $T$ denotes the transpose of matrix.

Optical pumping with light polarized along the $z$-direction will create the alignment 
\begin{equation}  \label{eq:m2eq}
\mathbf{m}_2^{\mrm{eq}} = \left(0,0, m_{2,0}^{\mrm{ini}}, 0, 0 \right)^T.
\end{equation}
If the light is instead polarized at an angle $\theta$ in the $zx$-plane, as illustrated in Fig.~\ref{fig:DRAM illustration}, we have 
\begin{equation} \label{eq:m2eqtheta}
\mathbf{m}_2^{\mrm{eq}}\left( \theta \right)  = \left(0,0, m_{2,0}^{\mrm{ini}} \left[1+3\cos(2\theta) \right]/4, 0, 0 \right)^T,
\end{equation}
which can be calculated by rotating the multipole given in Eq.~\ref{eq:m2eq} by an angle $-\theta$ around the $y$-axis and assuming that multipoles with $q\neq 0$ average to zero as we do not consider the pulsed optical pumping.

\subsection{Evolution of the atomic state}
We now calculate the evolution of the alignment multipoles in the presence of the magnetic field 
$B_0 \hat{z} + B_{\mathrm{rf}} \cos (\wRF t)\hat{x}$.
The unitary evolution is governed by the equation
\begin{equation}
\dot{\mbf{m}}_2 = 
\left(
-i \omega_L J_z^{(2)}
-i \Omega_{\text{rf}}  \cos (\wRF t) J_x^{(2)} 
\right) \mbf{m}_2 ,
\end{equation}
where 
$\omega_L = \gamma_F B_0/\hbar$ is the Larmor frequency, $\Omega_{\text{rf}} = \gamma_F B_{\text{rf}}/\hbar$ is the Rabi frequency of the RF field, and $\gamma_F$ is the gyromagnetic ratio.
The matrices $J_x^{(2)}$ and $J_z^{(2)}$ are the relevant representation of the angular momentum operators, which are written out in matrix form in Appendix~\ref{app:spinmatrices}.

We now define variables in the rotating frame with frequency $\wRF$
\begin{equation} \label{eq:rotframe}
\widetilde{m}_{\kappa,q}=e^{iq \wRF t} m_{\kappa,q},
\end{equation}
or equivalently written using vector notation
\begin{equation} \label{eq:mtilde}
\mathbf{\widetilde{m}}_2 = \mathbf{R}_z^{\left(2\right)}\left(-\wRF t \right) \mathbf{m}_2,
\end{equation}
where 
\begin{equation} \label{eq:Rthetaz}
\mathbf{R}_z^{\left(2\right)} \left(-\wRF t\right) 
= \exp \left( i \wRF t  J_z^{(2)} /\hbar \right)
\end{equation}
is the rotation matrix around the $z$ axis generated by the relevant angular momentum representation. 
%around the $z$ axis with the angle $-\wRF t$.
Assuming the rotating wave approximation, the rotating frame multipoles will evolve according to
\begin{equation}
\dot{\widetilde{\mbf{m}}}_2=
\left( i \delta J_z^{(2)} - i \frac{\Omega_{\text{rf}}}{2} J_x^{(2)} \right)
\mathbf{\widetilde{m}}_2
-\Gamma \mathbf{\widetilde{m}}_2 + \Gamma  \mbf{\widetilde{m}}_2^{\mrm{eq}},\label{eq:evolution}
\end{equation}
where $\delta=\omega_{\text{rf}}-\omega_L$ is the detuning of the RF field from the Larmor frequency. 
We have furthermore included terms describing optical pumping and ground state relaxation in the equation, and for simplicity assumed relaxation can be described by a single rate $\Gamma$.
Equation~\ref{eq:evolution} can be solved in the steady state where $\dot{\widetilde{\mbf{m}}}_2=0$ yielding the solution $\widetilde{\mbf{m}}_2^{\mathrm{ss}}$ which is explicitly written out in Appendix~\ref{app:multipoles}. 

\subsection{Polarization rotation signal}
\label{sec:PRsignal}
The observable signals depend on the multipoles 
$\mbf{m}'_2 $
that are defined with quantization axis in the light-polarization $\ehat$ direction. They are related to the lab frame multipoles $\mathbf{m}_2$ and one can calculate the other by applying a rotation with the angle $\theta$, i.e.
\begin{equation} \label{eq:mprime}
\mathbf{m}'_2 = \mathbf{R}_y^{\left(2\right)}\left(\theta\right) \mathbf{m}_2,
\end{equation}
where the rotation matrix around the $y$-axis with the angle $\theta$ is
\begin{equation} \label{eq:Rthetay}
\mathbf{R}_y^{\left(2\right)} \left(\theta\right) 
= \exp \left( - i \theta J_y^{(2)} /\hbar \right).
\end{equation}
Using the definition of the rotating frame, we furthermore have
\begin{equation} \label{eq:mprimeB}
\mathbf{m}'_2 = 
\mathbf{R}_y^{\left(2\right)}\left(\theta\right) 
\mathbf{R}_z^{\left(2\right)}
\left(+\wRF t\right)\widetilde{\mathbf{m}}_2 .
\end{equation}

We first consider the light polarization rotation signal which for light propagating in the $y$-direction can be calculated using the methods of \cite{meraki2023zerofield} as
\begin{equation}
    S^{\pr}(t) \propto \, 
    \left[ m'_{2,-1}(t) - m'_{2,1}(t) \right].
    \label{eq:meas_signal}
\end{equation}
Inserting the rotating frame steady state solutions given in App.~\ref{app:multipoles} into Eq.~\ref{eq:mprimeB}, and calculating the polarization rotation signal given by Eq.~\ref{eq:meas_signal}, we find that the signal has three contributions
\begin{equation}
    S^{\pr}(t) \propto S^{\pr}_0(t) + S^{\pr}_1(t)+ S^{\pr}_2(t),
\end{equation}
which are oscillating at frequencies $0$, $\rmdn{\ww}{rf}$ and $2\rmdn{\ww}{rf}$, respectively.
We can compactly write the signal as
\begin{align}
    S^{\pr}_0(t)& \propto h_0(\theta) A_0 ,\\
    S^{\pr}_1(t)&  \propto h_1(\theta)[D_1\cos(\rmdn{\ww}{rf}t)
          +A_1\sin(\rmdn{\ww}{rf}t)] ,\label{eq:PR_amp_1st} \\
    S^{\pr}_2(t)&  \propto h_2(\theta)[A_2\cos(2\rmdn{\ww}{rf}t)
    + D_2\sin(2\rmdn{\ww}{rf}t)] ,     \label{eq:PR_amp_2nd}
\end{align}
where the angular dependence takes the form 
\begin{subequations}
    \begin{align}
        h_0(\theta) &= \sin(2\theta)[1+3\cos(2\theta)] , \label{eq:PR_angular 0th Harmonic} \\
        h_1(\theta) &= \cos(2\theta)[1+3\cos(2\theta)] , \label{eq:PR_angular 1st Harmonic} \\
        h_2(\theta) &= \sin(2\theta)[1+3\cos(2\theta)]. \label{eq:PR_angular 2nd Harmonic}
    \end{align}
\end{subequations}
The in-phase and quadrature line shapes for the first harmonic signal take the form 
\begin{align}
\scalemath{0.95}{A_1 \left(\delta,\Omega_{\text{rf}} \right) =\frac{\sqrt{\frac{3}{2}} m_{2,0}^{\mathrm{ini}} \Gamma\Omega_{\text{rf}} \left[4 \left( \Gamma^2+4\delta^2\right)+\Omega_{\text{rf}}^2\right] 
}{
4(\Gamma^2+4\delta^2+\Omega_{\text{rf}}^2) \left[4 \left(\Gamma^2+\delta^2\right)+\Omega_{\text{rf}}^2 \right]
}}, \label{eq:PR-A1}  \\
\scalemath{0.95}{D_1(\delta,\Omega_{\text{rf}}) = -
\frac{ \sqrt{\frac{3}{2}}  m_{2,0}^{\mathrm{ini}} \delta \Omega_{\text{rf}} 
\left[2 \left(\Gamma^2+4\delta^2 \right)-\Omega_{\text{rf}}^2 \right]
}{
2(\Gamma^2+4\delta^2+\Omega_{\text{rf}}^2) \left[ 4 \left(\Gamma^2+\delta^2\right)+\Omega_{\text{rf}}^2 \right]
}.}
\label{eq:PR-D1} 
\end{align}
On resonance, the amplitude of the first harmonic signal $R_1=\sqrt{A_1^2+D_1^2}$ will be 
\begin{equation}
R_1 \left(\delta=0, \Omega_{\text{rf}} \right) =
\frac{\sqrt{\frac{3}{2}} m_{2,0}^{\mathrm{ini}} \Gamma\Omega_{\text{rf}} 
}{4\left(\Gamma^2+\Omega_{\text{rf}}^2\right) }. \label{eq:PR_1st vs RF}
\end{equation}
We see that the first harmonic signal is proportional to the RF intensity $\Omega_{\text{rf}}$ for small amplitude of $\Omega_{\text{rf}}$.
In the limit of low RF intensities, \eqnref{eq:PR-A1} and \eqnref{eq:PR-D1} reduce to
\begin{align}
A_1(\delta,\Omega_{\text{rf}} \approx 0) = &
\frac{1}{4} \sqrt{\frac{3}{2}}  m_{2,0}^{\mathrm{ini}}
\frac{\Gamma\Omega_{\text{rf}}}{\Gamma^2+\delta^2} ,
\label{eq:PR low RF absorptive} \\
D_1(\delta,\Omega_{\text{rf}}\approx 0) =&
-\frac{1}{4} \sqrt{\frac{3}{2}}  m_{2,0}^{\mathrm{ini}}
\frac{\delta\Omega_{\text{rf}}}{\Gamma^2+\delta^2} ,
\label{eq:PR low RF dispersive} \\
R_1(\delta,\Omega_{\text{rf}}\approx 0) = &
\frac{1}{4} \sqrt{\frac{3}{2}}  m_{2,0}^{\mathrm{ini}}
\frac{\Omega_{\text{rf}}}{\sqrt{\Gamma^2+\delta^2}} .
\label{eq:PR-R1limit}
\end{align}
We also calculate the signal at the second harmonic
\begin{align}
\scalemath{0.95}{A_2 \left(\delta,\Omega_{\text{rf}} \right) =
\frac{\sqrt{\frac{3}{2}}  m_{2,0}^{\mathrm{ini}}  \Omega_{\text{rf}}^2 
\left[4 \left( \Gamma^2-2\delta^2\right)+\Omega_{\text{rf}}^2\right]
}{
16(\Gamma^2+4\delta^2+\Omega_{\text{rf}}^2) \left[4 \left(\Gamma^2+\delta^2\right)+\Omega_{\text{rf}}^2 \right]
}}, \label{eq:PR-A2} \\
\scalemath{0.95}{D_2(\delta,\Omega_{\text{rf}}) = \frac{3 \sqrt{\frac{3}{2}}  m_{2,0}^{\mathrm{ini}} \Omega_{\text{rf}}^2  \Gamma \delta
}{
4 (\Gamma^2+4\delta^2+\Omega_{\text{rf}}^2) \left[ 4 \left(\Gamma^2+\delta^2\right)+\Omega_{\text{rf}}^2 \right]
}}.\label{eq:PR-D2}
\end{align}
On resonance, the second harmonic signal is
\begin{equation}
R_2 \left(\delta=0,\Omega_{\text{rf}} \right) =
\frac{\sqrt{\frac{3}{2}}  m_{2,0}^{\mathrm{ini}}  \Omega_{\text{rf}}^2 
}{
16\left(\Gamma^2+\Omega_{\text{rf}}^2\right) 
}. \label{eq:PR_2nd vs RF}
\end{equation}
For $\Omega_{\text{rf}}<<\Gamma$, the signal at the second harmonic is proportional to the square of the RF intensity $\Omega_{\text{rf}}^2$. In the limit of low RF intensities Eqs.~(\ref{eq:PR-A2}) and (\ref{eq:PR-D2}) simplify to 
\begin{align}
A_2 \left(\delta,\Omega_{\text{rf}}\approx 0\right) =
&\frac{\sqrt{\frac{3}{2}}  m_{2,0}^{\mathrm{ini}}  \Omega_{\text{rf}}^2
 \left( \Gamma^2-2\delta^2\right)
}{
16(\Gamma^2+4\delta^2) \left(\Gamma^2 +  \delta^2\right)
}, \label{eq:PR-A2limit} \\
D_2(\delta,\Omega_{\text{rf}}\approx 0) =
&\frac{3 \sqrt{\frac{3}{2}}  m_{2,0}^{\mathrm{ini}} \Omega_{\text{rf}}^2  \Gamma \delta
}{
16 (\Gamma^2+4\delta^2) \left(\Gamma^2+\delta^2\right)
} , \label{eq:PR-D2limit} \\
R_2(\delta,\Omega_{\text{rf}}\approx 0) = &
\frac{1}{16} \sqrt{\frac{3}{2}}  m_{2,0}^{\mathrm{ini}}
\frac{\Omega_{\text{rf}}^2}{\sqrt{\Gamma^4+5\Gamma^2 \delta^2 +4\delta^4}} .
\label{eq:PR-R2limit}
\end{align}
In our work we mainly focus on the signal at the first and second harmonics. However, for completeness, we also calculate the $A_0$ term which will give a DC rotation of the transmitted light polarization 
which in the limit of low RF intensities simplifies to 
\begin{equation}
A_0 \left(\delta,\Omega_{\text{rf}}\approx 0 \right) =
  \sqrt{\frac{3}{2}} m_{2,0}^{\mathrm{ini}}  
\left\{ 
\frac{1}{4} - \frac{3}{16} \frac{\Omega_{\text{rf}}^2}{ \left( \Gamma^2 + \delta^2 \right) }
\right\}.
 \label{eq:PRA0limit} 
\end{equation}

%%%%%%%%%%%%%%%%%%%%%%%%%%%%%%%%%%%%%%%%%%%%%%%%%%%%%%%%%%%%%%%%%%%%%%%%
\subsection{Absorption measurements}
We now consider the light absorption signal. 
In the limit of low absorption, the transmitted light intensity will be $I(L) = I(0)\exp \left(-\kappa_{\mathrm{abs}} L \right) \approx 
I(0)\left(1 -\kappa_{\mathrm{abs}} L \right)$, where $\kappa_{\mathrm{abs}}$ is the absorption coefficient, $L$ is the length of the vapour cell, and $I(0)$ and $I(L)$ are the light intensity before and after the vapour cell, respectively.
The absorption coefficient can be calculated as \cite{weis2006theory,meraki2023zerofield}
\begin{equation} \label{eq:kappa}
\kappa_{\mathrm{abs}} \propto m'_{0,0}-\chi m'_{2,0} ,
\end{equation}
where again the prime refers to the multipoles defined with the quantization axis along the light-polarization direction. Here $\chi$ depends on the hyperfine quantum number $F$.
The first term in Eq.~(\ref{eq:kappa}) is proportional to $m'_{0,0}=m_{0,0}$ and will give a constant offset. The second term,  proportional to $m'_{2,0}$, will also lead to a DC absorption as well as absorption oscillating at frequencies $\rmdn{\ww}{rf}$ and $2\rmdn{\ww}{rf}$.
We are mainly interested in the absorption at the first and second harmonics as those are measured in the experiments.
Following the same procedure as in Sec.~\ref{sec:PRsignal} and then calculating $m'_{2,0}$, we find
\begin{align}
    S^{\mathrm{abs}}_1(t) \propto &
    f_1(\theta)[X_1\cos(\rmdn{\ww}{rf}t)
          +Y_1\sin(\rmdn{\ww}{rf}t)] ,\\
    S^{\mathrm{abs}}_2(t) \propto&
    f_2(\theta)[X_2\cos(2\rmdn{\ww}{rf}t)
    + Y_2\sin(2\rmdn{\ww}{rf}t)] ,     
\end{align}
where the angular dependence takes the form 
\begin{subequations}
\begin{align}
    f_1(\theta)&=\sin(2\theta)[1+3\cos(2\theta)] = h_2(\theta) ,
    \label{eq-abs-f1}\\
    f_2(\theta)&=\left( 1- \cos(2\theta) \right) [1+3\cos(2\theta)], 
    \label{eq-abs-f2}
\end{align}
\end{subequations}
and where the coefficients $X_1$, $Y_1$, $X_2$, $Y_2$ can be related to ones derived previously for the polarization rotation signal as follows
\begin{subequations}
\begin{align}
X_1 =& -\frac{1}{4} \sqrt{6} D_1 , \label{eq-abs-X1}  \\
Y_1 =& -\frac{1}{4} \sqrt{6} A_1 ,  \label{eq-abs-Y1} \\
X_2 =& -\frac{1}{4} \sqrt{6} A_2 ,  \label{eq-abs-X2} \\
Y_2 =& -\frac{1}{4} \sqrt{6} D_2 . \label{eq-abs-Y2} 
\end{align} 
\end{subequations}
We see that the angular dependence for the absorption signal differs from that of the polarization rotation signal. However, the magnetic resonance lineshapes are the same for absorption and polarization rotation signals, up to a proportionality constant. Furthermore, our results for the absorption signals agree with what was calculated by Weis et al.\ \cite{weis2006theory}.
Note that Ref. \cite{weis2006theory} considers a rotating RF field while we consider an RF field applied along one spatial direction, which leads to a factor of two difference in the definitions of the Rabi frequencies of the RF field.
%%%%%%%%%%%%%%%%%%%%%%%%%%%%%%%%%%%%%%%%%%%%%%%%%%%%%%%%%%%%%%%%%%%%%%%% 
\section{Experimental Setup}\label{sec;experiment}
\begin{figure}
    \centering
    \includegraphics[width=0.485\textwidth]{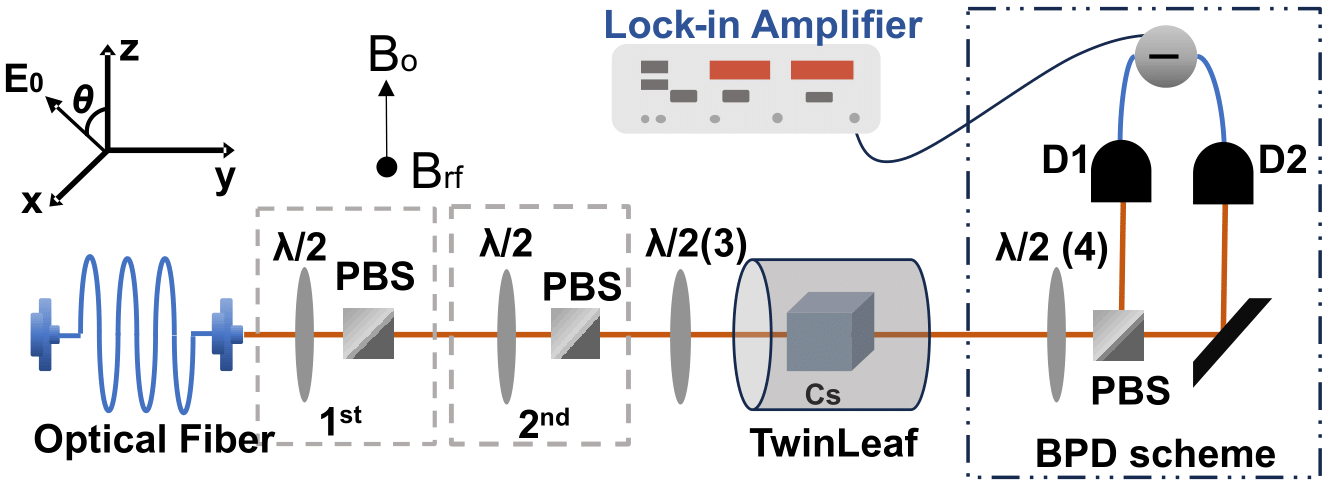}
    \caption{\RaggedRight The schematic of experimental setup for the double resonance alignment magnetometry. Here $\lambda/2$ = Half waveplate, PBS = Polarizing beamsplitter, Twinleaf = Magnetic shield, Cs = Caesium vapour cell, BPD = Balanced photodetector, D1 and D2 are photodetectors.} 
    \label{fig:experimental_setup}
\end{figure}

A schematic of the experimental setup for measuring magnetic resonance signals is shown in Fig.~\ref{fig:experimental_setup}. A polarization maintaining optical fiber outputs linearly polarized light ($\lambda= 894.6$ nm) propagating in the $y$-direction. The light frequency is resonant with D1 transition ($F=4 \rightarrow F^{'}=3$) of Cs atoms. The first pair of half waveplate and polarizing beamsplitter is used to minimise the polarization fluctuation of light whereas the second pair is used to control the incident light power. Another half waveplate $\lambda/2 (3)$ is utilized to rotate the polarization angle ($\theta$) of incident light in the $zx$-plane. 
The light, with an electric field amplitude $E_0$, is then passed through a cubic paraffin coated Cs cell that is placed inside a four layered mu-metal magnetic shield (Twinleaf MS-1). 
For $\theta= 0$, the electric field amplitude is $\mathbf{E}_0=E_0 \mathbf{\hat{z}}$ 
and for $\theta=90^{\circ}$, we have $ \mathbf{E}_0 = E_0 \mathbf{\hat{x}}$. A static magnetic field $\mathbf{B}_0$ is applied in the $z$-direction using a coil integrated in the Twinleaf magnetic shield. An oscillating weak RF magnetic field ($\mathbf{B}_{\text{rf}}= B_{1}\cos(2\pi f_{\text{rf}}t)\mathbf{\hat{x}}$) is applied in the $x$-direction using a homemade coil placed inside the magnetic shield. Here $f_{\mathrm{rf}}= \omega_{\mathrm{rf}}/2\pi$ is the RF frequency in Hz.
After the magnetic shield, 
a polarimetric technique based on balanced photodetection (BPD) using a half-wave plate $\lambda/2 (4)$, a polarizing beamsplitter and a balanced photodetector (Thorlabs PDB210A/M)
is employed to measure the polarization rotation of light. The output from the photodetector is then fed to a lock-in amplifier (Stanford Research Systems SR-830), which demodulates the signal for an adjusted harmonic. The in-phase (X) and quadrature (Y) signals are subsequently recorded using a data-acquisition card (Spectrum M2P.5932-X4) for further analysis. Note that for the absorption measurement, the BPD scheme is replaced by a single photodetector (Thorlabs PDA36A2).

\section{Results and Discussion}\label{sec;results and discussion}
\subsection{Experimental procedure}
\begin{figure}
    \centering
    \includegraphics[width=1\linewidth]{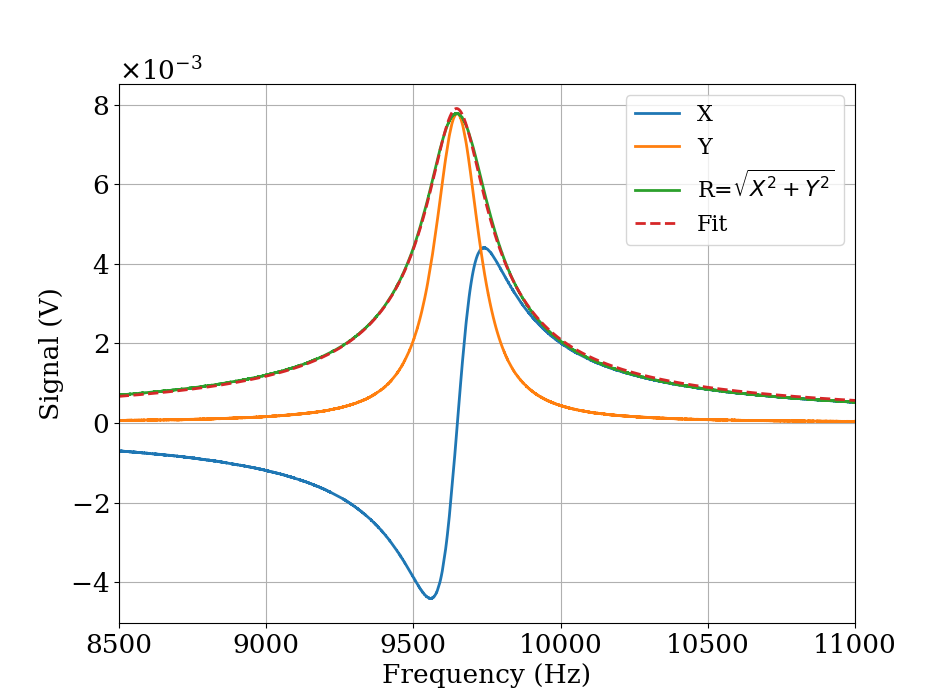}
    \captionsetup{justification=justified, singlelinecheck=false}
    \caption{\RaggedRight Example magnetic resonance signal for first harmonic observed through polarization rotation where light power is P$=6$ $\mu$W, lock-in sensitivity $= 10$ mV and incident light polarization angle $\theta = 0^\circ$. The dashed lines are curve fitted to the data R.
    }
    \label{fig:Resonance curve}
\end{figure}
\begin{figure*}
    \centering
    \begin{subfigure}[]{0.49\textwidth}
        \centering
        \includegraphics[width=1\linewidth]{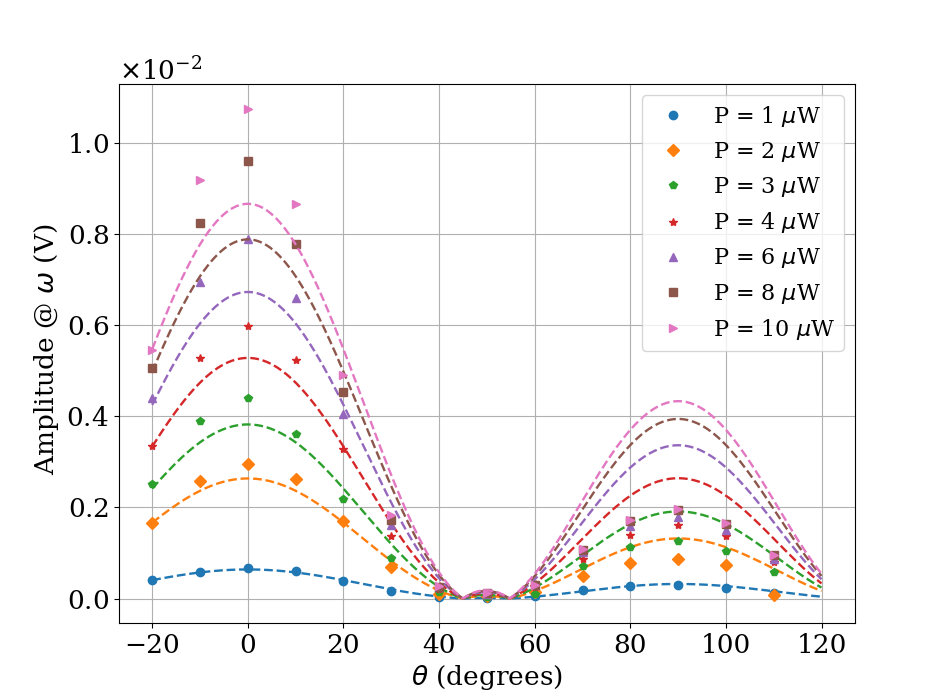}
\vspace{-0.5cm}
        \caption{}
        \label{fig:PR_1st harmonic}
    \end{subfigure}
        \hspace{-0.7cm}
    \begin{subfigure}[]{0.49\textwidth}
        \centering
        \includegraphics[width=1\linewidth]{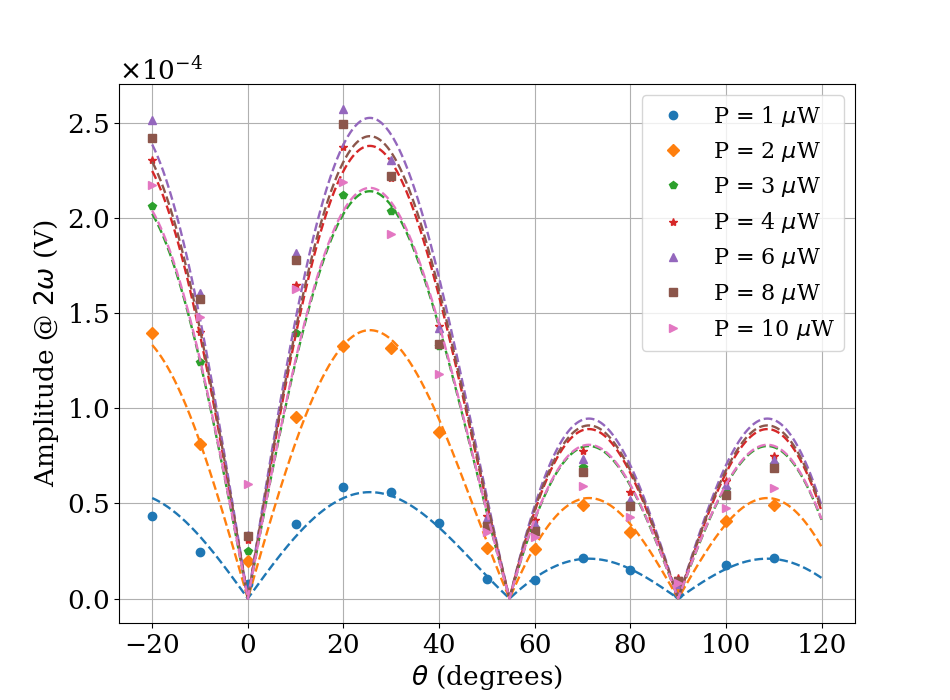}
\vspace{-0.5cm}

        \caption{}
        \label{fig:PR_2nd harmonic}
    \end{subfigure}
\vspace{-0.17cm}

    \begin{subfigure}[]{0.49\textwidth}
        \centering
        \includegraphics[width=1\linewidth]{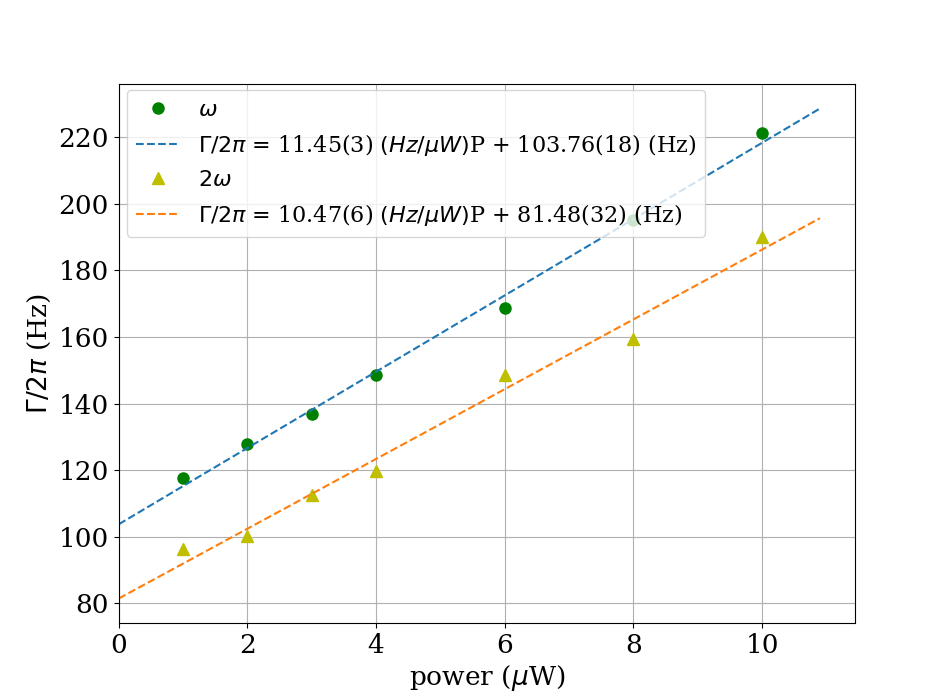}
        \caption{\label{fig:PR gamma_1st_n_2nd_harmonics}}
    \end{subfigure}
            \hspace{-0.7cm}
    \begin{subfigure}[]{0.49\textwidth}
        \centering
        \includegraphics[width=1\linewidth]{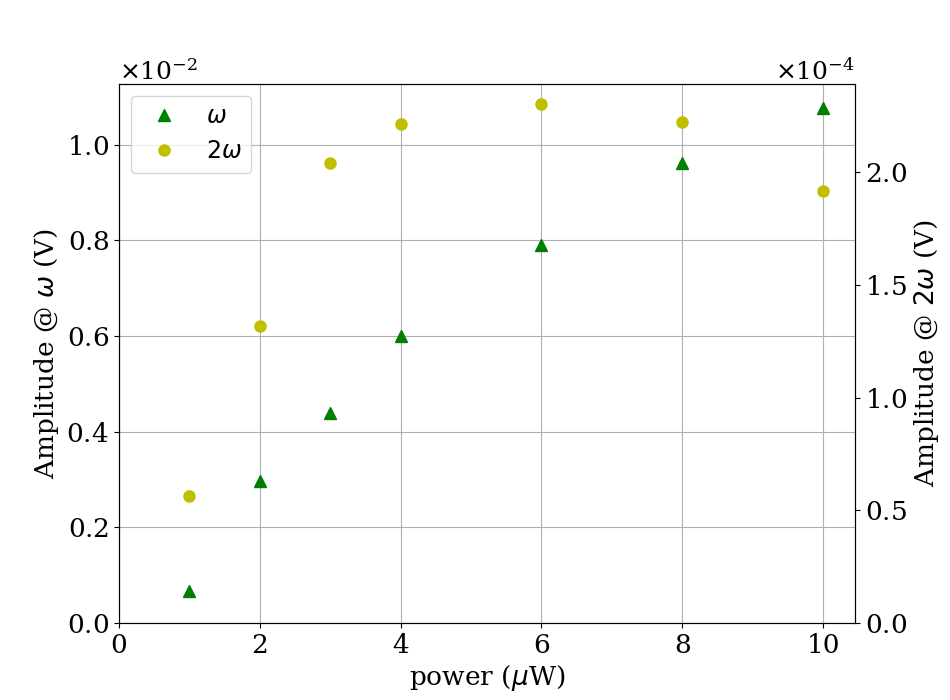}
        \caption{ \label{fig:PR_amplitude_1st_n_2nd_harmonics}}
    \end{subfigure}
        \captionsetup{justification=justified, singlelinecheck=false}
   \caption{\RaggedRight Polarization rotation measurement where on resonance amplitudes are plotted as a function of polarization angle of light for (a) first and (b) second harmonic. (c) Relaxation rate ($\Gamma/2\pi$) and (d) resonance signal amplitudes are depicted as a function of light power for first and second harmonic, respectively. The dashed lines are curve fit to the data.
   \label{fig:PR harmonics}}
\end{figure*}
We commence the discussion with the measurement of magnetic resonance signal. 
For these measurements, the static magnetic field strength is set to a value corresponding to a Larmor frequency of $f_L=\omega_{L}/2\pi\approx 9.5$ kHz. The frequency of the RF field is then swept from $8$--$12$ kHz in 5~s. The polarization rotation of light is detected with the balanced photodetector which is then demodulated by the lock-in amplifier to measure the in-phase and quadrature components. The time constant is set to $10$ ms for all measurements. At the onset of measurement, the phase of the lock-in is adjusted to get the dispersive curve crossing the zero at resonant frequency for the incident polarization angle of $\theta=0^\circ$. Subsequently, the lock-in phase was kept fixed throughout the measurements. A typical measurement of an in-phase (X) and quadrature (Y) signal for the first harmonic observed through polarization rotation is shown in Fig.~\ref{fig:Resonance curve}. Note that while analyzing data, the lock-in sensitivity (which determines the gain of the lock-in amplifier) is taken into account, such that the plotted voltage in Fig.~\ref{fig:Resonance curve} corresponds to the root-mean square amplitude of the oscillating input signal to the lock-in amplifier. The figure also shows the calculated amplitude $R=\sqrt{X^2+Y^2}$ which is fitted to Eq.~(\ref{eq:PR-R1limit}).

In order to capture the angular dependence of resonance signal at first and second harmonic, the half-wave plates ($\lambda/2 (3)$) and ($\lambda/2 (4)$) are mounted in motorized rotation stages (Thorlabs PRM$1$-Z$8$) with dc controllers (Thorlabs KDC-$101$) and the angle is changed with a step size of $10^{\circ}$ over a wider range of polarization angles ($-20^\circ$ to $110^\circ$). This experimental setting is repeated for different input light powers ($P = 1$--$10$ $\mu$W). The magnetic resonance spectra was recorded and curve fitted to the \eqnref{eq:PR-R1limit} and \eqnref{eq:PR-R2limit} for the first and second harmonic, respectively. This allows us to extract
the resonance frequency (where $f_{\mathrm{rf}}=f_L$)  in Hz, the relaxation rate $\Gamma/(2\pi)$ in Hz and the amplitude at resonance in Volts.
Note that we used a single relaxation rate $\Gamma$ for curve fitting which is contrary to the approach undertaken in Ref. \cite{di2006experimental}. Using an isotropic relaxation rate simplifies the analysis without undermining the theoretical framework. This experimental procedure is implemented for the measurement of both observables, i.e., intensity (absorption) and polarization rotation of light.

%%%%%%%%%%%%%%%%%%%%%%%%%%%%%%%%%%%
\subsection{Angular dependence of resonance spectra}

\begin{figure*}
    \centering
    \begin{subfigure}[]{0.49\textwidth}
    \includegraphics[width=1\linewidth]{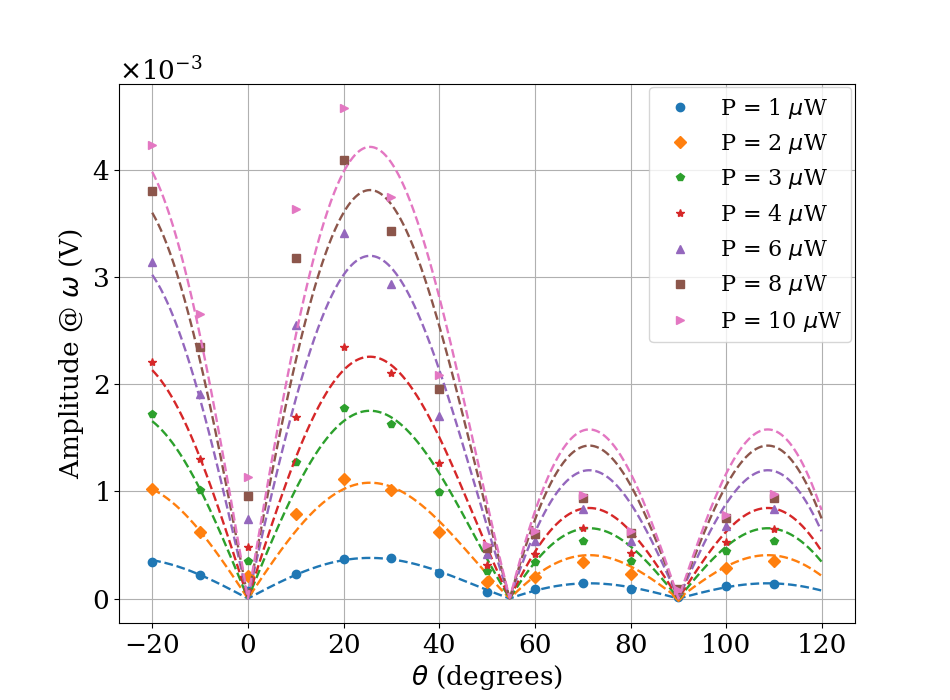}
\vspace{-0.5cm}

        \caption{}
        \label{fig:MR_absorption_1st}
    \end{subfigure}
            \hspace{-0.7cm}
    \begin{subfigure}[]{0.49\textwidth}
        \includegraphics[width=1\linewidth]{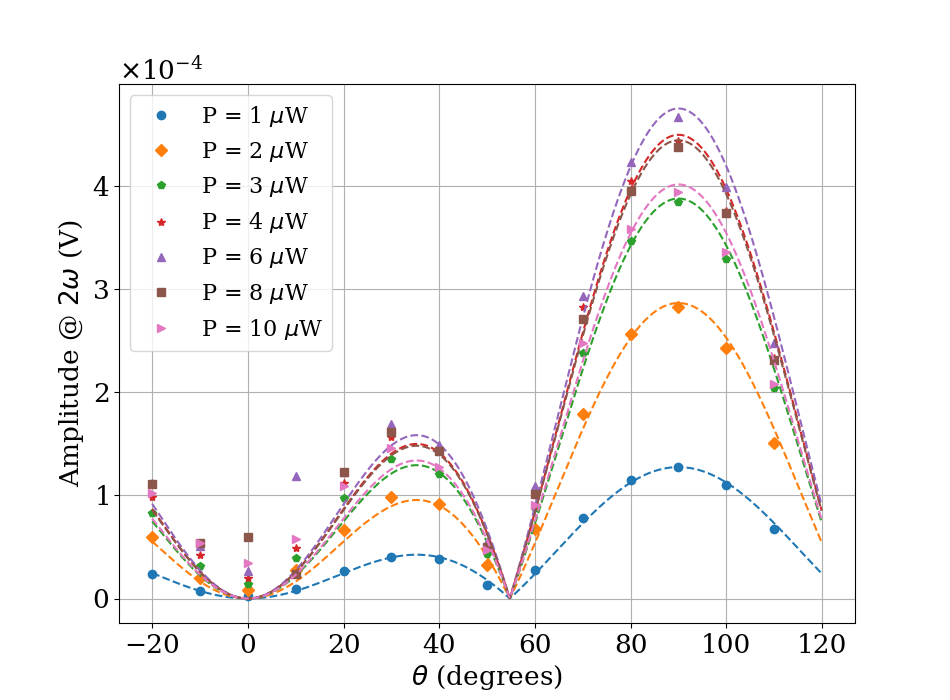}
\vspace{-0.5cm}

        \caption{}
        \label{fig:MR_absorption_2nd}
    \end{subfigure} 
\vspace{-0.175cm}

    \begin{subfigure}[]{0.49\textwidth}
        \centering
        \includegraphics[width=1\linewidth]{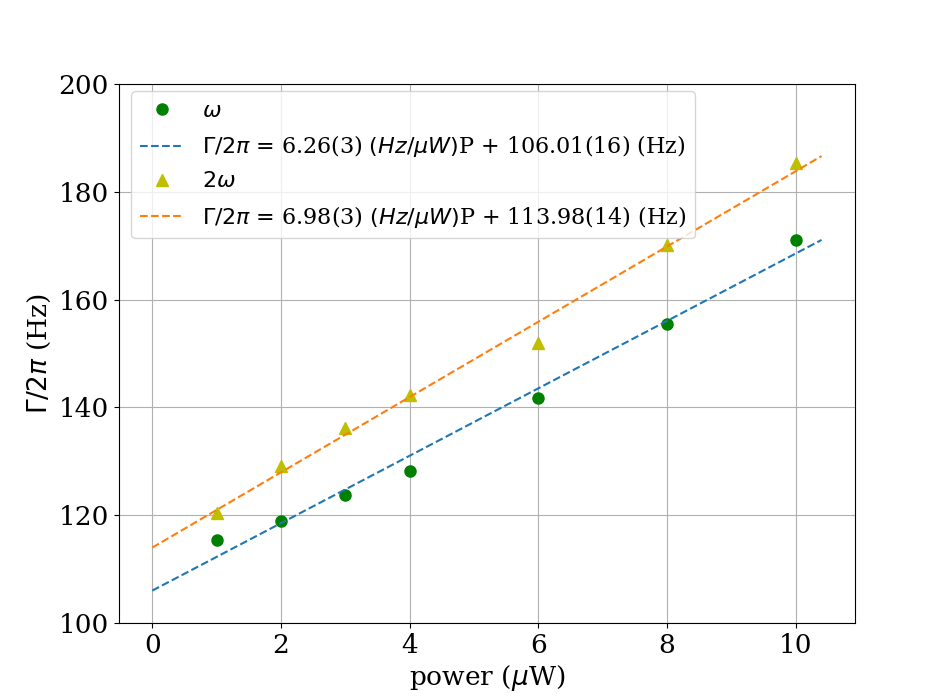}
\vspace{-0.5cm}

        \caption{}
        \label{fig:absorption_gammas}
    \end{subfigure}
            \hspace{-0.7cm}
    \begin{subfigure}[]{0.49\textwidth}
        \centering
        \includegraphics[width=1\linewidth]{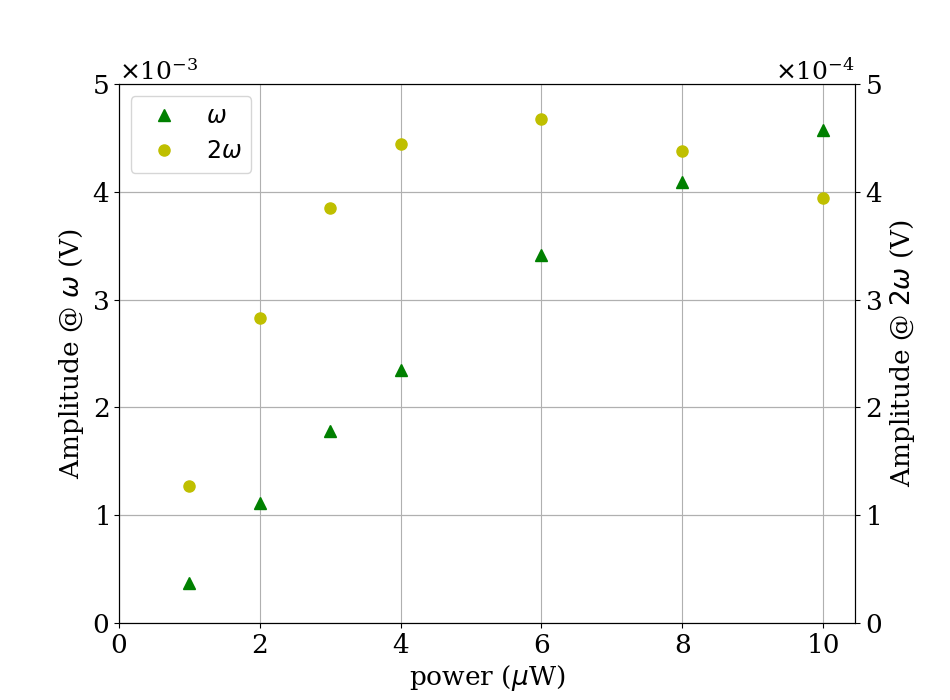}
\vspace{-0.5cm}

        \caption{}
        \label{fig:absorption_amplitudes}
    \end{subfigure}
        \captionsetup{justification=justified, singlelinecheck=false}
    \caption{\RaggedRight Absorption measurements where on resonance amplitude are plotted as a function of polarization angle of light for (a) first harmonic and (b) second harmonic whereas (c) relaxation rate ($\Gamma/2\pi$) and (d) on resonance amplitudes are plotted as a function of light power for first and second harmonic, respectively. The dashed lines are curve fit to the data.
    }
    \label{fig:MR_absorption harmonics}
\end{figure*}

Fig.~\ref{fig:PR_1st harmonic} and \ref{fig:PR_2nd harmonic} shows the results for first and second harmonic where magnetic resonance spectra is observed through polarization rotation of the transmitted light. The extracted fit parameter amplitude has an angular dependence, dictated by \eqnref{eq:PR_angular 1st Harmonic} and \eqnref{eq:PR_angular 2nd Harmonic} for first and second harmonic, respectively. It can be seen that experimental data fits reasonably well at the low powers of light. As the light power increases, experimental results tend to diverge from the theoretical description known to be valid only for weak probing.

Fig.~\ref{fig:PR gamma_1st_n_2nd_harmonics} illustrates the fit parameter for the relaxation rate $\Gamma/2\pi$ as a function of light power. The data fits to a linear function $\Gamma/2\pi= c \cdot P+ \Gamma_0/2\pi$ where $c$ is the slope with units of Hz/$\mu$W and $\Gamma_0/2\pi$ is the linewidth in Hz extrapolated to zero light power.
Fig.~\ref{fig:PR_amplitude_1st_n_2nd_harmonics} depicts the amplitude variation with input light power for a fixed value of polarization angle, i.e., $\theta=0^\circ$ for first and $\theta= 20^\circ$ for second harmonic. It can be seen that the amplitude for first harmonic increases monotonically with the light power. However, for second harmonic, amplitude reaches a maximum value and then follows a downward trend with further increase of light power. This provides an useful insight to optimize the controlling parameters such as light power, polarization angle to maximize the sensitivity of OPM to be used in DRAM geometry.

Similarly, the measurement of the absorption of light is carried out for various input light powers and incident light polarization angles. Fig.~\ref{fig:MR_absorption_1st} and \ref{fig:MR_absorption_2nd} shows the angular dependence of the amplitude of absorption signal for first and second harmonic, respectively. In this case, the data is curve fitted to the \eqnref{eq-abs-f1} and \eqnref{eq-abs-f2} for the relevant harmonic. It is noticeable that the fit function for the first harmonic agrees with the experimental data at low light power but less so for the higher light powers.  This is expected since the theoretical model is valid only for low light powers. Fig.~\ref{fig:absorption_gammas} and \ref{fig:absorption_amplitudes} illustrates the relaxation rate and amplitude of the resonance spectra as a function of input light power for first and second harmonic, respectively. It can be seen that relaxation rate again has a linear relationship with input light power and is curve fitted to a similar function described earlier ($\Gamma/2\pi=c \cdot P+\Gamma_0/2\pi$). 
The variation in amplitude of resonance signal is similar to what we observe in Fig.~\ref{fig:PR_amplitude_1st_n_2nd_harmonics}. However, it is noteworthy that amplitude value for first harmonic measured through polarization rotation is approximately $2.3$ times larger than amplitude measured in absorption signal. This underscores the enhanced sensitivity capability of polarization rotation based probing technique.
Moreover, when measuring absorption and polarization rotation, the magnetic resonance signal reaches zero at specific polarization angles ($\theta$) as dictated by the relevant equations (\ref{eq:PR_angular 1st Harmonic}, \ref{eq:PR_angular 2nd Harmonic}, \ref{eq-abs-f1}, \ref{eq-abs-f2}). However, we still observe small resonance signal at these angles. This small contribution can be ascribed to high light power.

%%%%%%%%%%%%%%
\subsection{Varying RF Amplitude}\label{sec;varying_RF_amplitude}
\begin{figure}
    \centering
    \includegraphics[width=0.50\textwidth]
    {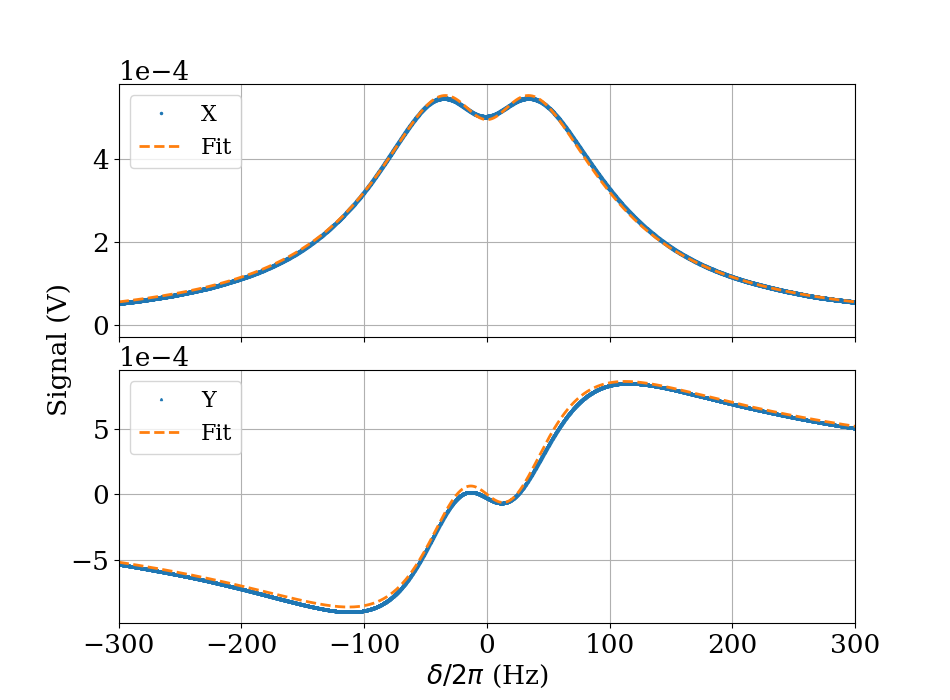}
    \captionsetup{justification=justified, singlelinecheck=false}
    \caption{\RaggedRight In-phase and quadrature magnetic resonance signal at higher amplitude of RF power plotted against the detuning $\delta$. The resonance lineshapes are measured through polarization rotation where P$= 10~ \mu$W, $\omega_L\approx 2\pi (9.5)$ kHz, $V_{\text{rf}}= 50$ mV and incident angle $\theta=40^\circ$. 
    }.
    \label{fig:High_RF_amplitude}
\end{figure}

\eqnref{eq:PR_1st vs RF} and \eqnref{eq:PR_2nd vs RF} capture the dependence of the magnetic resonance signal on the RF field amplitude for first and second harmonic, respectively. 
To validate the theory, we varied the RF field amplitude ($B_{\text{rf}}= g V_{\text{rf}}$) and measured the magnetic resonance spectra  for both harmonics through polarization rotation measurement. Here $g$ is the calibration factor with units (nT/V) and $V_{\text{rf}}$ is the controlling parameter for RF field amplitude. For the excitation coils, the calibration factor $(g)$ is determined and reads $210.8$ nT/V. For incident polarization angle $\theta = 0$, the light power is fixed at $10$ $\mu$W and Larmor frequency equals $f_L\approx 9.5$ kHz. Subsequently, the resonance data is recorded and curve fitted to \eqnref{eq:PR_1st vs RF} and \eqnref{eq:PR_2nd vs RF} for respective harmonic. The data is analyzed to extract the variable of interest such as amplitude (method as prescribed earlier in Section \ref{sec;results and discussion}). \\
\begin{figure}
    \centering
    \begin{subfigure}[b]{0.49\textwidth}
    \centering
    \includegraphics[width=1\linewidth]{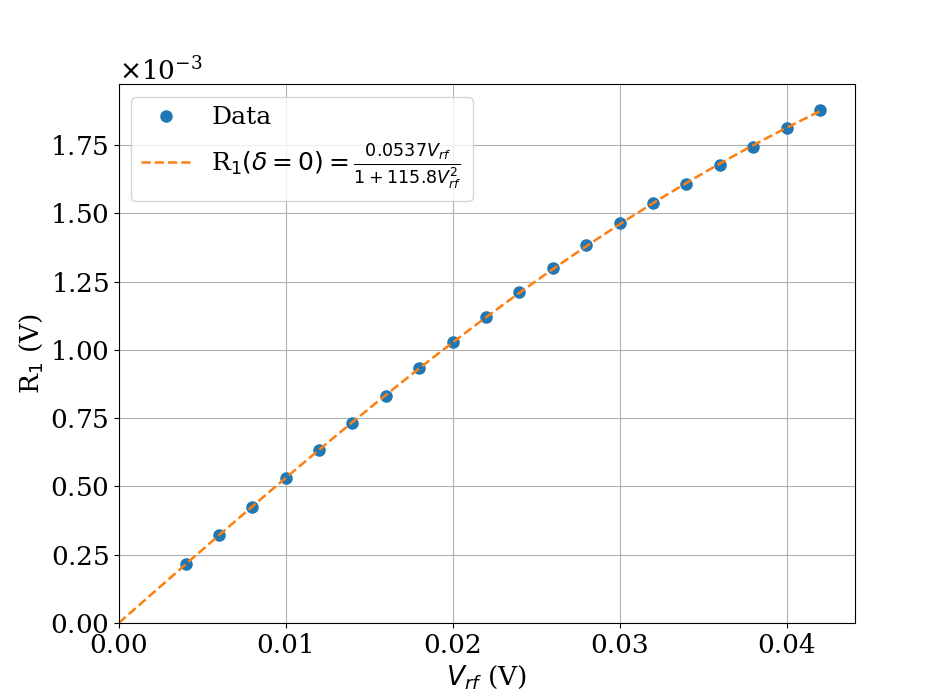}
\vspace{-0.5cm}

        \caption{}
        \label{fig:RF_amplitude_1st_harmonic}
    \end{subfigure}
\vspace{-0.53cm}

    \begin{subfigure}[b]{0.49\textwidth}
        \centering
        \includegraphics[width=1\linewidth]{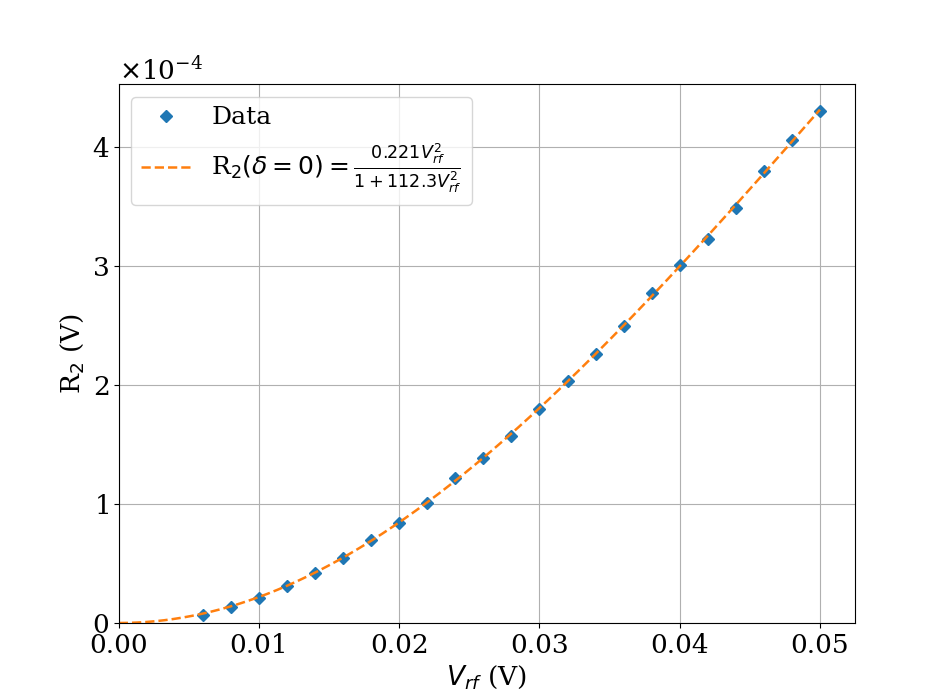}
\vspace{-0.5cm}

        \caption{}
        \label{fig:RF_amplitude_2nd_harmonic}
    \end{subfigure}
    \captionsetup{justification=justified, singlelinecheck=false}

    \caption{\RaggedRight On resonance amplitude of magnetic resonance spectra is plotted as a function of RF voltage to the excitation coils for (a) first harmonic and (b) second harmonic. The dashed lines represent curve fit to the data.}
    \label{fig:RF amplitude}
\end{figure}
Fig.~\ref{fig:RF_amplitude_1st_harmonic} and \ref{fig:RF_amplitude_2nd_harmonic} capture the variations in amplitude of magnetic resonance signal as a function of applied RF voltage ($V_{\text{rf}}$) for first and second harmonic, respectively. The functions fit the experimental data reasonably well. For low values of RF power, the amplitude exhibits a linear and quadratic relationship with RF power for first and second harmonic, respectively. This respective dependence tend to diverge at higher values of RF power, however \eqnref{eq:PR_1st vs RF} and \eqnref{eq:PR_2nd vs RF} completely describe the variations in amplitudes for arbitrary strength of RF power. 
It is worth mentioning here that in accordance with the theoretical model at higher magnitudes of RF power, we noted an additional feature in the magnetic resonance spectra which appears at the resonance frequency. 
Fig.~\ref{fig:High_RF_amplitude} illustrates the in-phase (X) and quadrature (Y) line spectra for high RF power. The in-phase and quadrature signals are fitted to the full expressions given by \eqnref{eq:PR-A1} and \eqnref{eq:PR-D1}, respectively. The results are in good agreement with the theoretical findings. 

\begin{figure}
    \centering
    \includegraphics[width=0.5\textwidth]{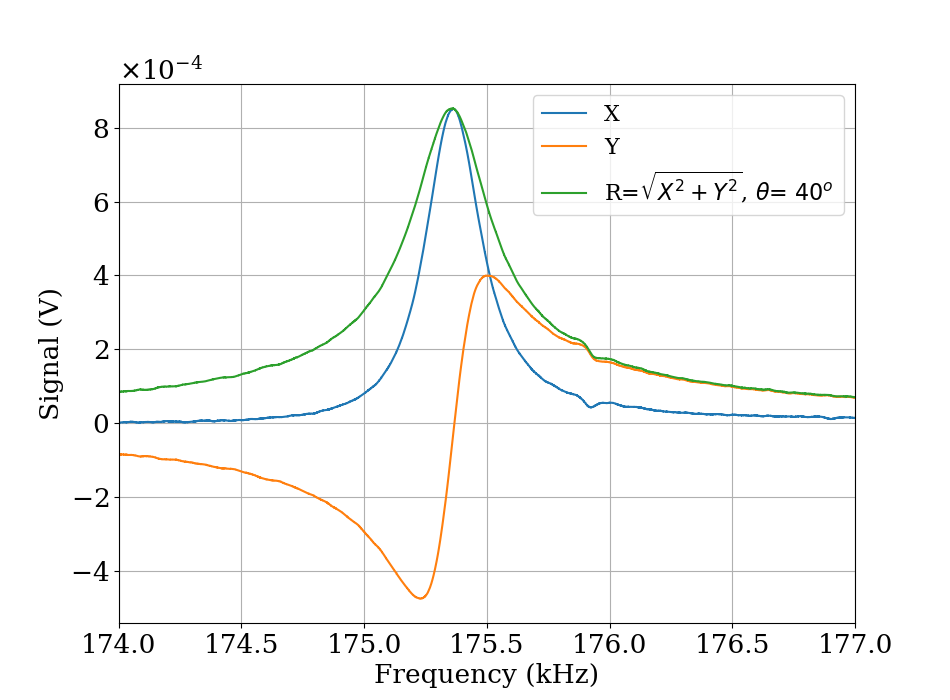}
    \caption{\RaggedRight Magnetic resonance spectra observed through polarization rotation of the light in the geo-magnetic field range which corresponds to a Larmor frequency of $\approx 175$ kHz with light power P$= 10$ $\mu$W.}
    \label{fig:MR at 175 kHz}
\end{figure}

It is pertinent to mention here that we also measured the magnetic resonance spectra in the earth's magnetic field range which gives the Larmor frequency of $\approx 175$ kHz. The corresponding magnetic field is applied through a TwinLeaf DC current source (CSU-$1000$). The observed lineshapes are presented in Fig.~\ref{fig:MR at 175 kHz} where incident light polarization makes an angle $40^\circ$ with the magnetic field $\mathbf{B}_0$. We observed a large resonance peak at $\approx 175.36$ kHz and one additional small peak at the frequency $\approx 175.92$ kHz with a separation of $\approx 560 $ Hz. While the large resonance is due to atoms in the $F=4$ hyperfine ground states, the small resonance is due to atoms in the $F=3$ hyperfine ground states. This can be seen from the different gyromagnetic ratio of Cs for the $F=3$ and $F=4$ state, i.e., $g_{F=3}/g_{F=4}=-1.0032$ \cite{steck}. This implies that if the $F=4$ resonance is at $175.36$ kHz then the $F=3$ resonance should be at  $-1.0032 \times 175.36$ kHz $=-175.92$ kHz. The minus sign signifies that the resonance lineshapes for two ground states will be $180^{\circ}$ out of phase with respect to each other and can be observed in Fig.~\ref{fig:MR at 175 kHz}. The $F=3$ resonance is small since the laser is $9.2$ GHz detuned from F$=3$ resonance. This peak is not observable in the resonance spectra at $9.5$ kHz since $9.5$ kHz $\times 1.0032 = 9.530$ KHz and falls inside the large resonance peak.  
\subsection{Sensitivity Measurement}\label{sec;sensitivity measurement}
\begin{figure*}
    \centering
   \begin{subfigure}{0.49\textwidth}
        \includegraphics[width=\linewidth]{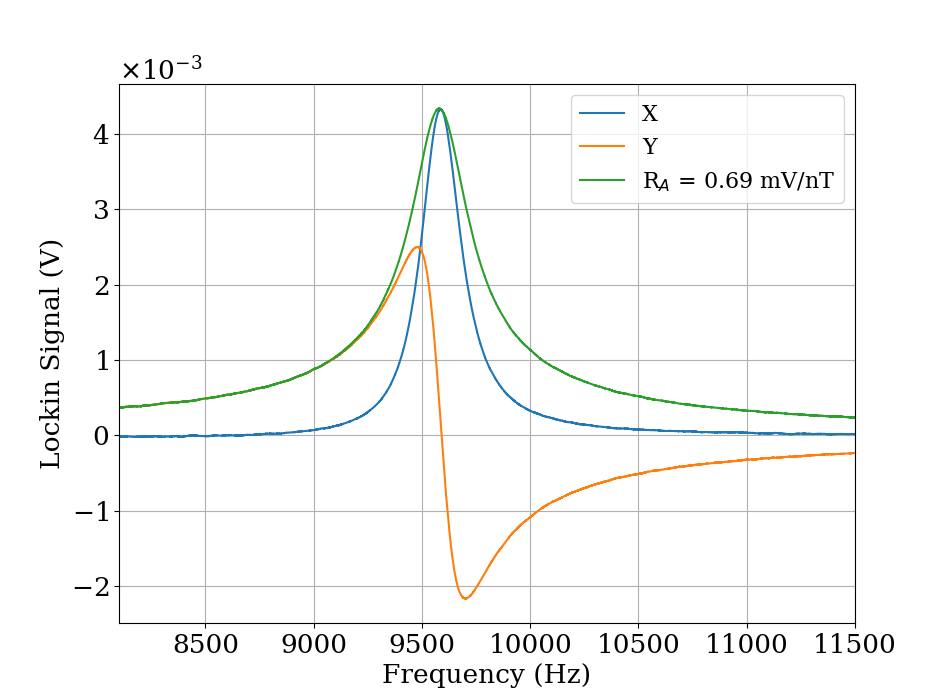}
        \caption{}
        \label{fig:MR_absorption}
    \end{subfigure}
            \hspace{-0.5cm}
    \begin{subfigure}{0.5\textwidth}
        \includegraphics[width=\linewidth]{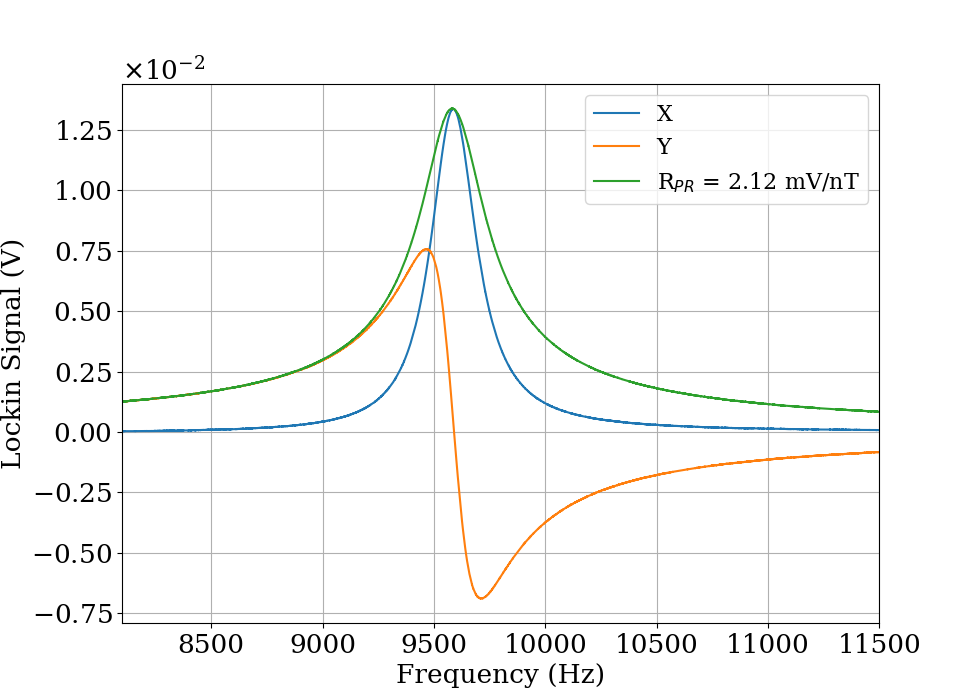}
        \caption{}
        \label{fig:MR_PR}
    \end{subfigure}
    
\vspace{-0.17cm} % Adjust the value as needed
    
    \begin{subfigure}{0.49\textwidth}
        \includegraphics[width=\linewidth]{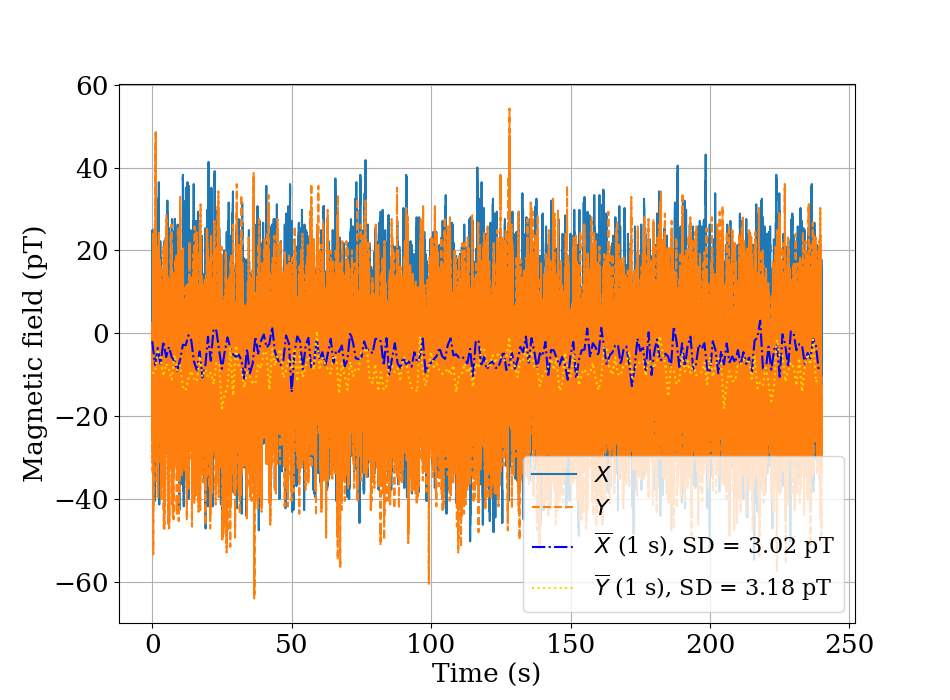}
        \caption{}
        \label{fig:absorption_sensitivity}
    \end{subfigure}
                \hspace{-0.5cm}
   \begin{subfigure}{0.5\textwidth}
        \includegraphics[width=\linewidth]{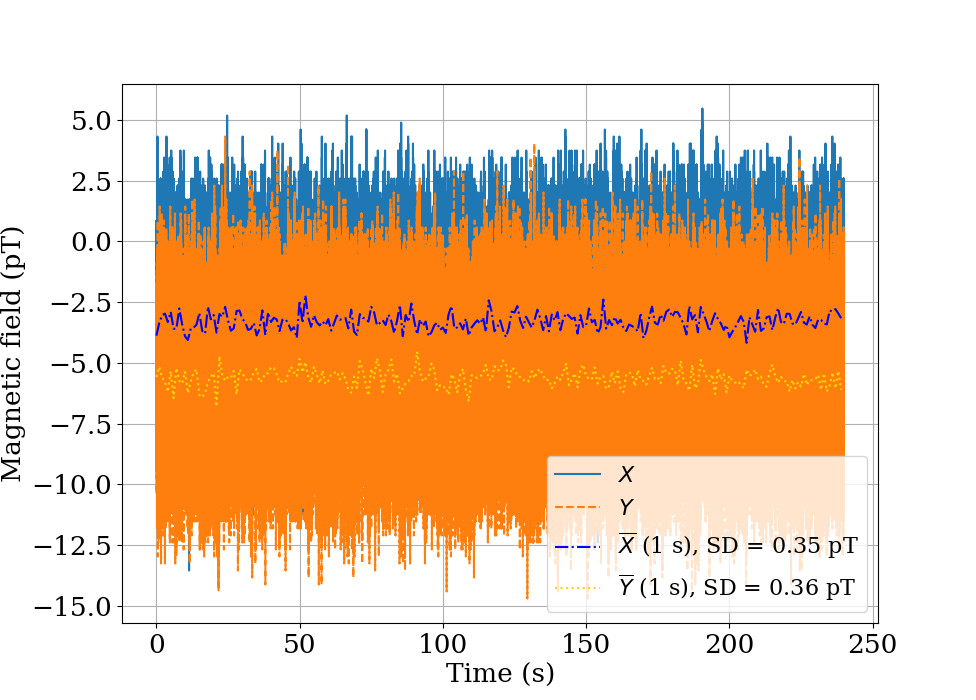}
        \caption{}
        \label{fig:PR_sensitivity}
    \end{subfigure}
    
    \caption{\RaggedRight First harmonic signal for in-phase, quadrature, and signal R measured through (a) absorption and (b) polarization rotation measurement, whereas (c) and (d) show their respective sensitivities of the magnetic field. The light power is $10$ $\mu$W and incident polarization angle is $\theta= 25^\circ$ and $\theta= 0^\circ$ for absorption and polarization rotation, respectively.
    }
    \label{fig:sensitivities}
\end{figure*}

Figure~\ref{fig:MR_absorption} and \ref{fig:MR_PR} shows an RF frequency sweep with $f_L =9.5$~kHz for absorption and polarization rotation measurement, respectively.  
For an applied oscillating field of $6.324$~nT, the peak value of the resonance spectra is extracted and is subsequently divided by the RF magnetic field to give a conversion factor between the lock-in amplifier readout and the corresponding RF field amplitudes. For absorption measurement, it reads $R_{A}= (0.69(2)$~mV/nT) whereas for polarization signal, it is given by $R_{PR}=(2.12(3)$~mV/nT). We see that the polarization rotation signal is approximately a factor of three larger than for the absorption measurement. Note that the transimpedance gains of the two detectors are almost identical ($4.75\cdot 10^5$~V/A and $5.0\cdot 10^5$~V/A). The RF frequency is then fixed to the Larmor frequency, i.e.,~$f_{\mathrm{rf}}=f_L$, and a 4 minute time trace is obtained. This measurement is repeated with RF amplitude set to zero, i.e.,~``RF off''. The recorded data is then analysed to estimate the sensitivity of OPM for oscillating magnetic fields by the method prescribed in \cite{rushton2023alignment} where standard deviation of $240\times 1$ s averaged segments is calculated.
Fig.~\ref{fig:absorption_sensitivity} shows the sensitivity to small oscillating magnetic fields for absorption detection mode and reads ~3.02 pT/$\sqrt{\text{Hz}}$ for $X$ and 3.18~pT/$\sqrt{\text{Hz}}$ for $Y$. 
Similarly, sensitivity measurement from polarization rotation is performed and results are presented in Fig.~\ref{fig:PR_sensitivity}. The magnetic field sensitivity for the $X$ and $Y$ are $\approx 0.35$ pT/$\sqrt{\text{Hz}}$ and $\approx 0.36$ pT/$\sqrt{\text{Hz}}$, respectively. Note that the sensitivity to magnetic field ($\delta B_{\text{meas}}$) measured through polarization rotation is approximately an order of magnitude better than absorption measurement. \\
In order to identify the intrinsic noise sources in the system, the light hitting the photodetector (Thorlabs PDB210A/M for the polarization rotation measurements and Thorlabs PDA36A2 for the absorption measurements) is blocked and another time trace is recorded (data not shown). The noise with the light blocked is 0.25~pT/$\sqrt{\text{Hz}}$ (for both $X$ and $Y$) for the polarization measurement and  1.34~pT/$\sqrt{\text{Hz}}$ for the absorption measurement.
This noise contribution is  due to the electronic noise of the photodetector and data acquisition card.\\
To augment the analysis, we calculated the shot noise estimation for both detection modalities and details are presented in Appendix \ref{shot noise estimation}. The estimated shot noise values are $0.80$ pT/$\sqrt{\text{Hz}}$ and $0.26$~pT/$\sqrt{\text{Hz}}$ for single and balanced photodetector, respectively. The fundamental limit to the sensitivity of an OPM is given by spin projection noise which is given by
\begin{equation}
    \delta \mathrm{B_{proj}} = \frac{2 \hbar}{g_F \mu_B \sqrt{nVT_2}},
\end{equation}
where $\mu_B$ is the Bohr magneton, $g_F$ = 1/4 for the $F = 4$ Cs ground state, $n \approx 2.2 \times 10^{16} \mathrm{m^{-3}} $ is the number density of Cs atoms at room temperature (18.5$^{\circ}$C) \cite{rushton2023alignment} and $V = (5 \mathrm{mm})^3$ is the volume of our vapour cell. $T_2$ is the transverse relaxation time and is calculated as
$T_2=1/2\pi\Gamma \approx 4.5$ ms for polarization rotation and $T_2 = 5.8$ ms for absorption measurement. Here we used the $\Gamma$ values from Fig.~\ref{fig:PR gamma_1st_n_2nd_harmonics} and Fig.~\ref{fig:absorption_gammas}. The estimated spin projection noise for the absorption and polarization rotation measurements are $23$ fT/$\sqrt{\text{Hz}}$ and $26$ fT/$\sqrt{\text{Hz}}$, respectively. 
The measured magnetic field sensitivities and electronic noise, along with the estimated shot noise and  projection noise are presented in Table.~\ref{tab:magnetic_field_errors}.
\begin{table}
\centering
\vspace{0.1cm}
\begin{tabular}{lll||ll}
\hline
 (pT/$\sqrt{\text{Hz}}$)& $\delta B_{\text{meas}} $ &   $\delta B_{\text{Elec}}$ & $\delta B_{\text{Shot}}$ & $\delta B_{\text{proj}}$  \\
\hline
Absorption & 3.0 & 1.34 & 0.80 & 0.023 \\
Polarization & 0.35 & 0.25 & 0.26 & 0.026  \\
\hline
\end{tabular}
\caption{\RaggedRight Magnetic field sensitivities for the alignment based paraffin coated Cs cell OPM where $\delta B_{\text{meas}}$ is the measured sensitivity to magnetic field and $\delta B_{\text{Elec}}$ is the electronic noise whereas $\delta B_{\text{shot}}$ is the estimated shot noise and $\delta B_{\text{proj}}$ is the estimated projection noise.}
\label{tab:magnetic_field_errors}
\end{table}
Moreover, we measured the light noise power spectral density with balanced and single photodetectors, see App.~\ref{PSD} (Fig.~\ref{fig:PR_spectrum} and \ref{fig:Absorption_spectrum}). The average of the data in frequency range of interest ($9$--$10.5$ kHz) is taken and plotted as a function of optical power. It can be seen that for polarization measurements (Fig.~\ref{fig:PR_linear}), using a balanced photodetector, one can achieve a shot-noise limited measurement (as seen from the linear dependence on light power) whereas for absorption measurements (Fig.~\ref{fig:absorption_polynom}) with a single detector, one will measure classical/technical light noise (as seen from the quadratic dependence on light power). We emphasize that the polarization rotation measurement yields better magnetic field sensitivity, approximately an order of magnitude in our experimental setting, than the absorption measurement. 
We note that the magnitude of spin projection noise is approximately identical for both techniques since the optical pumping and evolution of atomic ensemble remains the same. Moreover, the signal amplitude and the sensitivity to magnetic field can be improved by heating the vapour cell.

\section{Conclusion}
In this study, we present theoretical description for polarized atomic ensemble for double resonance alignment magnetometer. We derive the analytical expressions for first and second harmonic signals where magnetic resonance spectra is accessed through polarization rotation and absorption of transmitted light through polarized atomic (Cs) vapours. The magnetic resonance spectra is investigated for various input parameters such as RF-field amplitude, input light power, angle between the magnetic field and the linear light polarization. The experimental results demonstrate a good agreement with theoretical findings. Furthermore, the sensitivity measurements are performed with both detection modalities, i.e., absorption and polarization rotation. It is found that polarization rotation offers better sensitivity, approximately an order of magnitude, than absorption measurement. Moreover if experimental setting is limited by shot noise, the sensitivity through polarization rotation is better by a factor of $3$. 
Our results motivate the use of polarization rotation detection modality  for prototype alignment based magnetometers, given the better sensitivity despite the added complexity in the experimental setup.
\begin{acknowledgements}
This work was supported by 
the QuantERA grant C’MON-QSENS! funded by the Engineering and Physical Sciences Research Council (EPSRC) Grant No.~EP/T027126/1, 
 the EPSRC Grant No.~EP/Y005260/1, 
the Novo Nordisk Foundation (Grant No.~NNF20OC0064182). Project C’MON-QSENS! is supported by the National Science Centre (2019/32/Z/ST2/00026), Poland under QuantERA, which has received funding from the European Union's Horizon 2020 research and innovation programme under grant agreement no. 731473.
\end{acknowledgements}
\appendix
\section{Angular Momentum Matrices} 
\label{app:spinmatrices}

\begin{equation}
J_x^{(2)} = \hbar
\begin{pmatrix} 
0 & 1 & 0 & 0 & 0 \\
1 & 0 & \sqrt{\frac{3}{2}} & 0 & 0 \\
0 & \sqrt{\frac{3}{2}} & 0 & \sqrt{\frac{3}{2}} & 0 \\
0 & 0 & \sqrt{\frac{3}{2}} & 0 & 1 \\
0 & 0 & 0 & 1 & 0 
\end{pmatrix},
\end{equation}
\vspace{-0.2cm}
\begin{equation}
J_y^{(2)} = \hbar
\begin{pmatrix}
0 & i & 0 & 0 & 0 \\
-i & 0 & i\sqrt{\frac{3}{2}} & 0 & 0 \\
0 & -i\sqrt{\frac{3}{2}} & 0 & i\sqrt{\frac{3}{2}} & 0 \\
0 & 0 & -i\sqrt{\frac{3}{2}} & 0 & i \\
0 & 0 & 0 & -i & 0
\end{pmatrix},
\end{equation}
\begin{equation}
J_z^{(2)} = \hbar
\begin{pmatrix} 
-2 & 0 & 0 & 0 & 0 \\
0 & -1 & 0 & 0 & 0 \\
0 & 0 & 0 & 0 & 0 \\
0 & 0 & 0 & 1 & 0 \\
0 & 0 & 0 & 0 & 2 
\end{pmatrix}.
\end{equation}
%\end{widetext}
\vspace{0.5cm}
\section{Steady state rotating frame multipoles} \label{app:multipoles}
The steady state multipole $\widetilde{\mathbf{m}}_2^{\mathrm{ss}}$ in the rotating frame has the following components
\begin{widetext}
\begin{align}
\widetilde{m}_{2,-2}^{\mathrm{ss}} = &
\frac{
-\sqrt{\frac{3}{2}} m_{2,0}^{\mathrm{eq}}(\theta) \Omega_{\text{rf}}^2
\left[ 
4 \left(\Gamma - \ii \delta \right) 
\left( \Gamma - 2 \ii \delta \right) 
+ \Omega_{\text{rf}}^2
\right]
}{
4 \left( \Gamma^2 +4 \delta^2 +\Omega_{\text{rf}}^2 \right) 
\left[ 4 \left(\Gamma^2+\delta^2 \right)+\Omega_{\text{rf}}^2\right] 
} , \label{eq:msteady1} \\
\widetilde{m}_{2,-1}^{\mathrm{ss}} = &
\frac{
- \ii \sqrt{\frac{3}{2}} m_{2,0}^{\mathrm{eq}}(\theta) 
\left( \Gamma + 2 \ii \delta \right) \Omega_{\text{rf}}
\left[ 
4 \left(\Gamma - \ii \delta \right) 
\left( \Gamma - 2 \ii \delta \right) 
+ \Omega_{\text{rf}}^2
\right]
}{
2 \left( \Gamma^2 +4 \delta^2 +\Omega_{\text{rf}}^2 \right) 
\left[ 4 \left(\Gamma^2+\delta^2 \right)+\Omega_{\text{rf}}^2\right] 
} , \label{eq:msteady2} \\
\widetilde{m}_{2,0}^{\mathrm{ss}} = &
\frac{
m_{2,0}^{\mathrm{eq}}(\theta) 
\left[
16 \left( \Gamma^2 + \delta^2 \right)
\left( \Gamma^2 + 4\delta^2 \right)
+
8 \left( \Gamma^2 - 2  \delta^2 \right) \Omega_{\text{rf}}^2
+ \Omega_{\text{rf}}^4
\right]
}{
4 \left( \Gamma^2 +4 \delta^2 +\Omega_{\text{rf}}^2 \right) 
\left[ 4 \left(\Gamma^2+\delta^2 \right)+\Omega_{\text{rf}}^2\right] 
} , \label{eq:msteady3} \\
\widetilde{m}_{2,1}^{\mathrm{ss}} = &
\frac{
- \ii \sqrt{\frac{3}{2}} m_{2,0}^{\mathrm{eq}}(\theta) 
\left( \Gamma - 2 \ii \delta \right) \Omega_{\text{rf}}
\left[ 
4 \left(\Gamma + \ii \delta \right) 
\left( \Gamma + 2 \ii \delta \right) 
+ \Omega_{\text{rf}}^2
\right]
}{
2 \left( \Gamma^2 +4 \delta^2 +\Omega_{\text{rf}}^2 \right) 
\left[ 4 \left(\Gamma^2+\delta^2 \right)+\Omega_{\text{rf}}^2\right] 
} , \label{eq:msteady4} \\
\widetilde{m}_{2,2}^{\mathrm{ss}} = &
\frac{
-\sqrt{\frac{3}{2}} m_{2,0}^{\mathrm{eq}}(\theta) \Omega_{\text{rf}}^2
\left[ 
4 \left(\Gamma + \ii \delta \right) 
\left( \Gamma + 2 \ii \delta \right) 
+ \Omega_{\text{rf}}^2
\right]
}{
4 \left( \Gamma^2 +4 \delta^2 +\Omega_{\text{rf}}^2 \right) 
\left[ 4 \left(\Gamma^2+\delta^2 \right)+\Omega_{\text{rf}}^2\right] 
} .
\label{eq:msteady5}
\end{align}
\end{widetext}
%\newpage
\section{Shot Noise Estimation}\label{shot noise estimation}
The  shot noise for a photodetector is given by $ \Delta i/\sqrt{\Delta f}= \sqrt{2eI}$ where $e$ is the electronic charge, $ I= R(\lambda) P$ is the average current from the photodetector with $R(\lambda)$ as responsivity and P is the incident light power on the detector. For $10$~$\mu$W light measured before the vapour cell, and an estimated $73\%$ transmission of light through the vapour cell, the power on the single detector will be $7.3$~$\mu$W. This gives a shot noise estimation in A/$\sqrt{\text{Hz}}$. Since the detector output is in volts and we also measure voltage signal with the lock-in, it is more useful to write the formula for shot noise in terms of V/$\sqrt{\text{Hz}}$ which reads as follows
\begin{equation}
\frac{\Delta V}{\sqrt{\Delta f}}= \frac{\Delta i}{\sqrt{\Delta f}} \big(G \times S.F\big), 
\end{equation}
where G is the trans-impedance amplifier (TIA) gain of the photodetector and $S.F$ is the scale factor which equals $\approx 1$ due to high impedance of lock-in input channel. For absorption technique, the estimated shot noise at P $=10~\mu$W then reads $\approx 5.6 \times 10^{-3}~ V/\sqrt{\text{Hz}}$ where we have also taken into account the internal gain of lock-in amplifier.  The value of shot noise is then subsequently divided by the conversion factor RA to
convert the shot noise (V /$\sqrt{\text{Hz}}$) in units of magnetic field and reads $0.80$ pT/$\sqrt{\text{Hz}}$.
Similar calculations were performed for the balanced photodetector by taking into account the respective gain and responsivity of the detector, the transmission through the vapour cell and a further estimated transmission of $92\%$ through the polarizing beamsplitter. Subsequently, the value of shot noise is divided by the conversion factor $R_{\text{PR}}$ and the estimated shot noise reads $0.26$~pT/$\sqrt{\text{Hz}}$. 

\section{Power Spectral Density Measurement}\label{PSD}
For different input light power, time traces were obtained. The power spectral density (PSD) of each trace is obtained. The PSD trace data is split into 50 time traces and subsequently, average of power spectral densities is taken and plotted.
%\counterwithin{figure}{section}
%\setcounter{figure}{0}
\begin{figure*}
\centering
   \begin{subfigure}{0.48\textwidth}
        \includegraphics[width=0.9\linewidth]{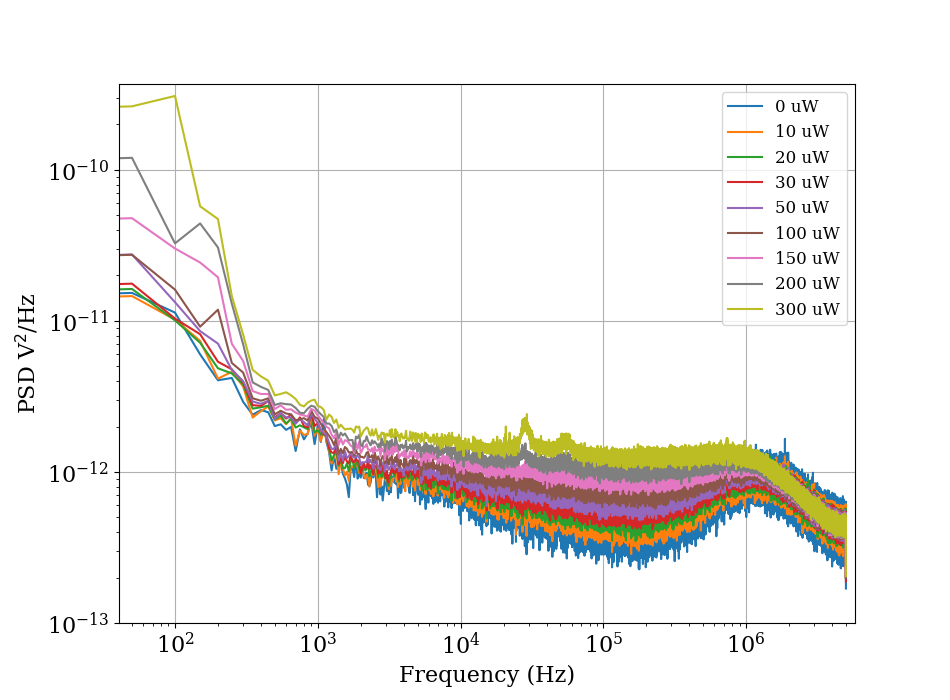}
        \caption{}
        \label{fig:PR_spectrum}
    \end{subfigure}
    \hspace{-1.1cm}
    \begin{subfigure}{0.48\textwidth}
        \includegraphics[width=0.9\linewidth]{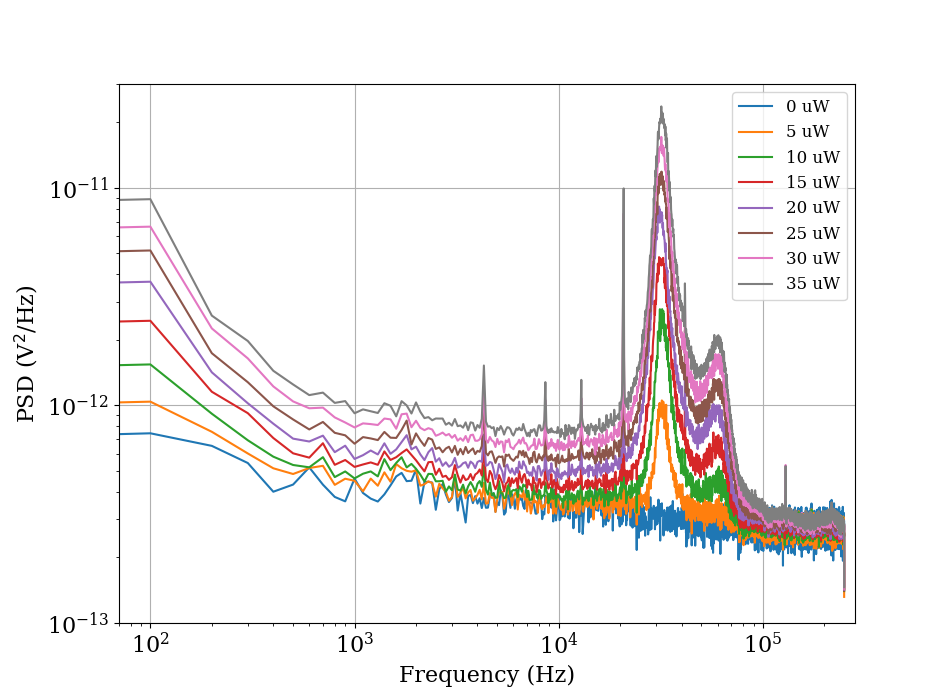}
        \caption{}
        \label{fig:Absorption_spectrum}
    \end{subfigure}
 
\vspace{-0.18cm} % Adjust the value as needed
    
   \begin{subfigure}{0.48\textwidth}
       \includegraphics[width=0.9\linewidth]{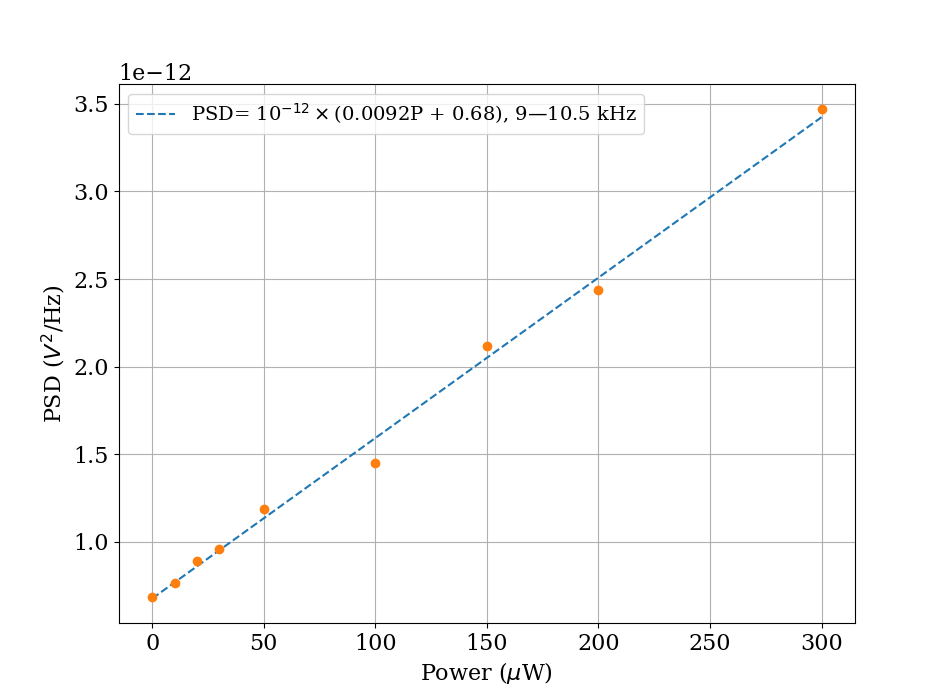}
       \caption{}
       \label{fig:PR_linear}
    \end{subfigure}
    \hspace{-1.1cm}
   \begin{subfigure}{0.48\textwidth}
        \includegraphics[width=0.9\linewidth]{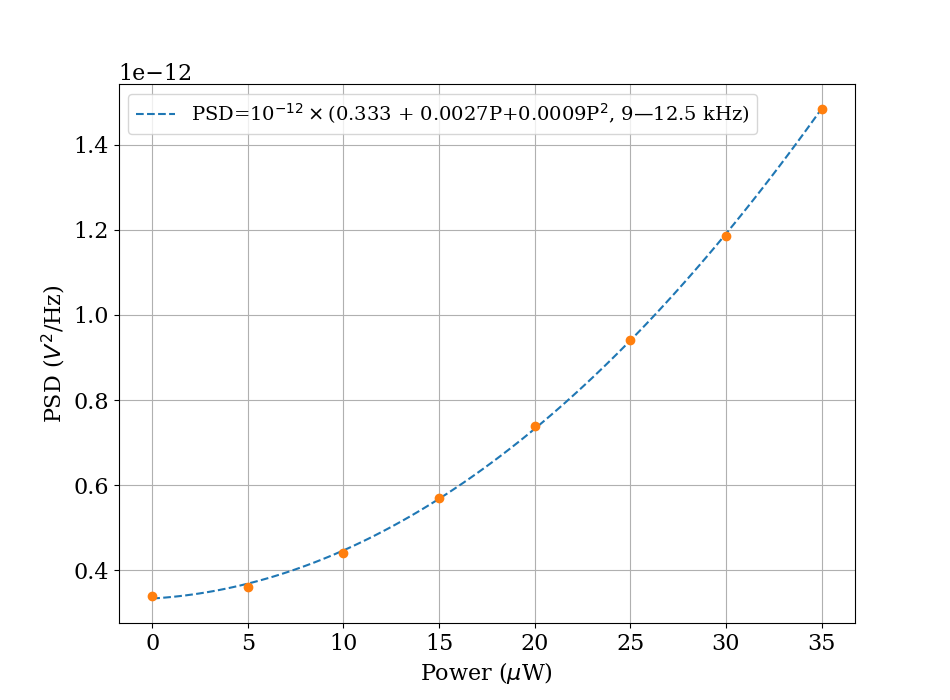}
        \caption{}
        \label{fig:absorption_polynom}
    \end{subfigure}
    
    \caption{\RaggedRight  PSD with (a) balanced detector and (b) single photodetector. The average PSD in the frequency range $9$--$10.5$ kHz is plotted as a function of optical power for (c) balanced and (d) single photodetector.}
    \label{fig:}
\end{figure*}

%\clearpage
%\bibliographystyle{unsrt}
%\bibliography{References}

\end{document}